\documentclass[pra,twocolumn,aps,letterpaper, preprintnumbers,superscriptaddress]{revtex4}

\usepackage{hyperref}
\usepackage{verbatim}
\usepackage{amsmath}
\usepackage{latexsym}
\usepackage{revsymb}
\usepackage{yfonts}
\usepackage{ifthen}
\usepackage{natbib}
\usepackage{amsfonts}
\usepackage{amsmath}
\usepackage{amssymb}
\usepackage{amsthm}
\usepackage{graphicx}
\usepackage{bm}
\usepackage{bbm}
\usepackage{epsfig,color,amssymb}
\usepackage{subfigure}
\usepackage{amsfonts}
\usepackage{amscd}
\usepackage{amsmath}
\usepackage{multirow}
\usepackage{chemarrow}
\usepackage{dcolumn}
\usepackage{bm}
\usepackage{graphicx}
\usepackage{enumerate}
\usepackage{epsfig}
\usepackage{subfigure}
\usepackage{xcolor}
\usepackage{multirow}
\usepackage{ulem}
\usepackage{braket}
\usepackage{comment}
\usepackage{enumitem}
\usepackage{amsthm}
\usepackage{algorithm}
\usepackage{algpseudocode}
\usepackage{overpic,color}
\renewcommand{\emph}[1]{\textit{#1}}

\begin{document}

\title{Automated machine learning for secure key rate in discrete-modulated continuous-variable quantum key distribution}
	
	\author{Zhi-Ping Liu}
	\affiliation{National Laboratory of Solid State Microstructures, School of Physics and Collaborative Innovation Center of Advanced Microstructures, Nanjing University, Nanjing 210093, China}
	
	\author{Min-Gang Zhou}
	\affiliation{National Laboratory of Solid State Microstructures, School of Physics and Collaborative Innovation Center of Advanced Microstructures, Nanjing University, Nanjing 210093, China}
	
	\author{Wen-Bo Liu}
	\affiliation{National Laboratory of Solid State Microstructures, School of Physics and Collaborative Innovation Center of Advanced Microstructures, Nanjing University, Nanjing 210093, China}
	
	\author{Wen-Bo Liu}
	\affiliation{National Laboratory of Solid State Microstructures, School of Physics and Collaborative Innovation Center of Advanced Microstructures, Nanjing University, Nanjing 210093, China}
	
	\author{Jie Gu}
	\affiliation{National Laboratory of Solid State Microstructures, School of Physics and Collaborative Innovation Center of Advanced Microstructures, Nanjing University, Nanjing 210093, China}

	\author{Hua-Lei Yin}
	\email{hlyin@nju.edu.cn}
	\affiliation{National Laboratory of Solid State Microstructures, School of Physics and Collaborative Innovation Center of Advanced Microstructures, Nanjing University, Nanjing 210093, China}
	
	\author{Zeng-Bing Chen}
	\email{zbchen@nju.edu.cn}
	\affiliation{National Laboratory of Solid State Microstructures, School of Physics and Collaborative Innovation Center of Advanced Microstructures, Nanjing University, Nanjing 210093, China}
	
	\date{\today}

\begin{abstract}Continuous-variable quantum key distribution (CV QKD) with discrete modulation has attracted increasing attention due to its experimental simplicity,  lower-cost implementation and compatibility with classical optical communication.  Correspondingly, some novel numerical methods have been proposed to analyze the security of  these protocols against collective attacks, which promotes key rates over one hundred kilometers of fiber distance. However, numerical methods are limited by their calculation time and resource consumption, for which they cannot play more roles on mobile platforms in quantum networks. To improve this issue, a neural network model predicting key rates in nearly real time has been proposed previously. Here, we go further and show a neural network model combined with Bayesian optimization. This model automatically designs the best architecture of neural network computing key rates in real time. We demonstrate our model with two variants of  CV QKD protocols with quaternary modulation. The results show high reliability with secure probability as high as $99.15\%-99.59\%$, considerable tightness and high efficiency with speedup of approximately $10^7$ in both cases. This inspiring model enables  the real-time computation of unstructured quantum key distribution protocols' key rate more automatically and efficiently, which has met the growing needs of implementing QKD protocols on moving platforms. 
\end{abstract}

\maketitle

\section{Introduction}

In recent decades, machine learning (ML) has gained impressive breakthroughs that deeply impact both industry and academia, including  autonomous driving~\cite{grigorescu2020survey,levinson2011towards}, natural language processing~\cite{deng2013new,young2018recent}, protein structure prediction~\cite{jumper2021highly} and even proving mathematical conjectures~\cite{davies2021advancing}. ML aims to recognize patterns in data, especially multidimensional data, and generalize them to new instances, which contributes to automating tasks and reveals hidden patterns beyond humans intuition. This modern information-processing technology also benefits solving intractable quantum tasks, since quantum tasks are usually counterintuitive and involve high dimensions. Several significant advances have been made by applying ML to quantum physics, from classifying quantum states~\cite{gao2018experimental,yang2019experimental,ahmed2021classification}, quantum control~\cite{bukov2018reinforcement,lumino2018experimental,niu2019universal} to quantum metrology~\cite{hentschel2011efficient}.

Quantum key distribution (QKD) enables unconditional security between two legitimate users (Alice and Bob) against any eavesdropper called Eve~\cite{bennett1984Quantum,ekert1991quantum}, which is guaranteed by quantum mechanics laws~\cite{shor2000simple}. According to different detection methods, QKD is currently divided into two categories: discrete-variable (DV) QKD~\cite{lo2014secure,gisin2002quantum} and continuous-variable (CV) QKD~\cite{grosshans2002continuous,lance2005no,huang2015high,yin2019phase}. Between these two categories, CV QKD has unique edges on a higher secret key rate and  excellent compatibility with standard communication components~\cite{fossier2009field,huang2016field,jouguet2012field}, which enables CV QKD to be competitive at a metropolitan distance~\cite{pirandola2020advances}. To enhance the practicality of CV QKD, several works introduce machine learning-based methodologies to the CV QKD area, such as developing a novel CV QKD scheme~\cite{jin2021key,liao2020multi}, parameter prediction~\cite{liu2018integrating} and detecting quantum attacks~\cite{mao2020detecting}.

CV QKD protocols with discrete modulation have attracted increasing attention for decades. Its appealing advantages include easier experimental implementation and higher error-correction efficiencies which promote CV QKD over longer distances~\cite{xu2020secure, leverrier2009unconditional, leverrier2011continuous, zhao2009asymptotic}. These properties bring potential advantages in large-scale deployment in quantum-secured networks~\cite{simon2017towards}. However, the security analysis of discrete-modulated CV QKD protocols is more complicated owing to the lack of symmetry~\cite{coles2016numerical}. Recently, some novel numerical approaches~\cite{lin2019asymptotic,winick2018reliable} have been proposed to analyze the security of discrete-modulation protocols against collective attacks, where key rate calculation involves minimizing a convex function over all eavesdropping attacks that are consistent with the experimental data. These numerical approaches achieve much higher key rates over significantly longer distances compared with previous security analyses. Based on these numerical approaches, a neural network model was presented to quickly predict the secure key rate of discrete-modulated CV QKD with high reliability (secure probability as high as $99.2\%$). This neural network model learns the mapping between input parameters and key rates from datasets generated by numerical methods, which supports the computation of secure key rates in real time~\cite{zhou2021machine}. However, the mapping complexity between input parameters and key rates depends on the solving complexity of discrete-modulated protocols' key rates through numerical approaches~\cite{hu2021robust}. Selecting architectures and hyperparameters plays a critical role in the performance of a neural network. Therefore, to learn different mappings from different protocols, the architectures of neural networks and the corresponding hyperparameters should be adjusted carefully by humans, which comes at a great price~\cite{yu2020hyper}.

Here, we propose a more flexible and automatic neural network model combined with Bayesian optimization~\cite{shahriari2015taking}, which maintains extremely high reliability and efficiency and reduces complicated manual adjustment. Our method is universal for a variety of unstructured QKD protocols that lack analytical tools and rely on numerical methods. We apply our model to two variants of discrete-modulated CV QKD protocols and acquire high secure key rates with considerable tightness in both cases. We then compare the time consumption of our model with the numerical method proposed in Ref.~\cite{lin2019asymptotic}, which shows a great speedup of approximately $10^7$. 

This paper is organized as follows. In section ~\ref{section 2}, we introduce the numerical method for CV QKD with discrete modulation proposed in Ref.~\cite{lin2019asymptotic}, and we rely on it to collect a dataset to train and test the model. In section ~\ref{section 3}, we introduce more details about the Bayesian optimization used in this paper. In section ~\ref{section 4}, we demonstrate all the main results of this paper. Section  ~\ref{section 5} provides a discussion and concludes this paper.

\section{Numerical method for CV QKD with discrete modulation}
\label{section 2}
In this work, we apply the model in two discrete-modulated CV QKD protocols with different detection techniques to demonstrate the generalizability of our model. One is the quadrature phase-shift-keying (QPSK) heterodyne detection protocol~\cite{lin2019asymptotic}, and the other is  an improved QPSK homodyne detection protocol~\cite{PRXQuantum.2.040334}. To collect a dataset for training neural networks, we generate secure key rates of both protocols by applying the same numerical method~\cite{lin2019asymptotic,hu2021robust}. In the following, we briefly introduce how computing key rates can be transformed into a relevant convex objective function for numerical optimization. A more detailed description can be found in Ref.\cite{lin2019asymptotic}.

Here, we consider a CV QKD protocol with quaternary modulation that involves two parties: a sender Alice and a receiver Bob. During each time in an iteration of N rounds, Alice randomly prepares one of the four coherent states $\left|\alpha_{k}\right\rangle =\left || \alpha |e^{i (2 k \pi/4+\pi/ 4)}\right \rangle$, where $k \in \{0,1,2,3\}$, and sends it to Bob via an untrusted quantum channel. Then, Bob uses either homodyne or heterodyne detection to estimate k. The secret key rate under collective attacks in the asymptotic limit is given by the following expression according to the Devetak-Winter formula~\cite{devetak2005distillation}

\begin{equation}
R^{\infty}=p_{\text {pass }}[\min _{\rho \in \mathbf{S}} H(\mathbf{Z} \mid E)-\delta_{\mathrm{EC}}]
\end{equation}
where $H\left(\mathbf{Z} \mid E\right)$ is conditional von Neumann entropy, which describes the uncertainty of the string $\mathbf{Z}$ in Eve’s view. Eve’s maximal knowledge of Bob’s string  $\mathbf{Z}$ requires the minimum uncertainty of  $\mathbf{Z}$ under a certain density matrix $\rho$. Therefore, we need to find the optimum $\rho^*$ in feasible domain  $\mathbf{S}$ to minimize $H\left(\mathbf{Z} \mid E\right)$, $p_{\text {pass }}$ is the sifting probability, and $\delta_{\mathrm{EC}}$ is the actual amount of information leakage per signal in the error-correction step. To turn this problem into a convex optimization problem, the above expression can be reformulated as 

\begin{equation}
R^{\infty}=\min _{\rho_{A B} \in \mathrm{S}} D\left(\mathcal{G}\left(\rho_{A B}\right) \| \mathcal{Z}\left[\mathcal{G}\left(\rho_{A B}\right)\right]\right)-p_{\text {pass }} \delta_{\mathrm{EC}}
\end{equation}
in which $D(\rho \| \sigma)=\operatorname{Tr}\left(\rho \log _{2} \rho\right)-\operatorname{Tr}\left(\rho \log _{2} \sigma\right)$. As shown in Ref.\cite{winick2018reliable}, $\mathcal{G}$ is a completely positive and trace nonincreasing map that describes the postprocessing of different quadratures. $\mathcal{Z}$ is a pinching quantum channel that reads out the key information. 

Since the term $p_{\text {pass }} \delta_{\mathrm{EC}}$ in formula $\left(2\right)$ is easy to compute, we can only consider the following relevant optimization problem:

\begin{align}\label{op}
&\operatorname{minimize}  D\left(\mathcal{G}\left(\rho_{A B}\right) \| \mathcal{Z}\left[\mathcal{G}\left(\rho_{A B}\right)\right]\right) \\
\text{subject to } \nonumber \\
&\operatorname{Tr}\left[\rho_{A B}\left(|k\rangle\left\langle\left. k\right|_{A} \otimes \hat{q}\right)\right]=p_{k}\langle\hat{q}\rangle_{k}\right. \\
&\operatorname{Tr}\left[\rho_{A B}\left(|k\rangle\left\langle\left. k\right|_{A} \otimes \hat{p}\right)\right]=p_{k}\langle\hat{p}\rangle_{k}\right. \\
&\operatorname{Tr}\left[\rho_{A B}\left(|k\rangle\left\langle\left. k\right|_{A} \otimes \hat{n}\right)\right]=p_{k}\langle\hat{n}\rangle_{k}\right. \\
&\operatorname{Tr}[\rho_{A B}(|k\rangle \left\langle k\right|_{A} \otimes \hat{d})]=p_{k}\langle\hat{d}\rangle_{k} \\
&\operatorname{Tr}\left[\rho_{A B}\right]=1 \\
&\rho_{A B} \geq 0\\
&\operatorname{Tr}_{B}\left[\rho_{A B}\right]=\sum_{i, j=0}^{3} \sqrt{p_{i} p_{j}}\left\langle\alpha_{j} \mid \alpha_{i}\right\rangle|i\rangle\left\langle\left. j\right|_{A}\right.
%\end{split}
%\end{equation}
\end{align}
where $k \in \left\{0,1,2,3\right\}$, $\langle\hat{q}\rangle_{k}$, $\langle\hat{p}\rangle_{k}$, $\langle\hat{n}\rangle_{k}$ and $\langle\hat{d}\rangle_{k}$ denote the expectation values of corresponding operators when Bob measures states labeled by $k$. These expectation values can be obtained through homodyne or heterodyne measurements. These first four constraints come from experimental outcomes. The next two constraints are natural requirements since $\rho_{AB}$ is a density matrix. The last constraint on the partial trace of system B comes from the fact that the quantum channel cannot influence system A of Alice. We can handle the above density matrix and operators in finite dimensions $N_c$ after imposing the photon-number cutoff assumption on this optimization problem~\cite{ghorai2019asymptotic,lin2019asymptotic}. Then, this problem can be solved numerically. Eventually, we solve this minimization problem by the numerical method proposed in Ref.\cite{winick2018reliable}. The specific implementation of this numerical method in our work can be found in Ref.\cite{PRXQuantum.2.040334}. This method involves two steps:

1. Find a solution that is close to optimal, which gives an upper bound on the key rate. 

2. Convert this upper bound to a lower bound on the key rate by considering its dual problem.  

\begin{figure}[ht]
\begin{center}
\vspace{0.2cm}
\begin{overpic}[width=9.0cm]{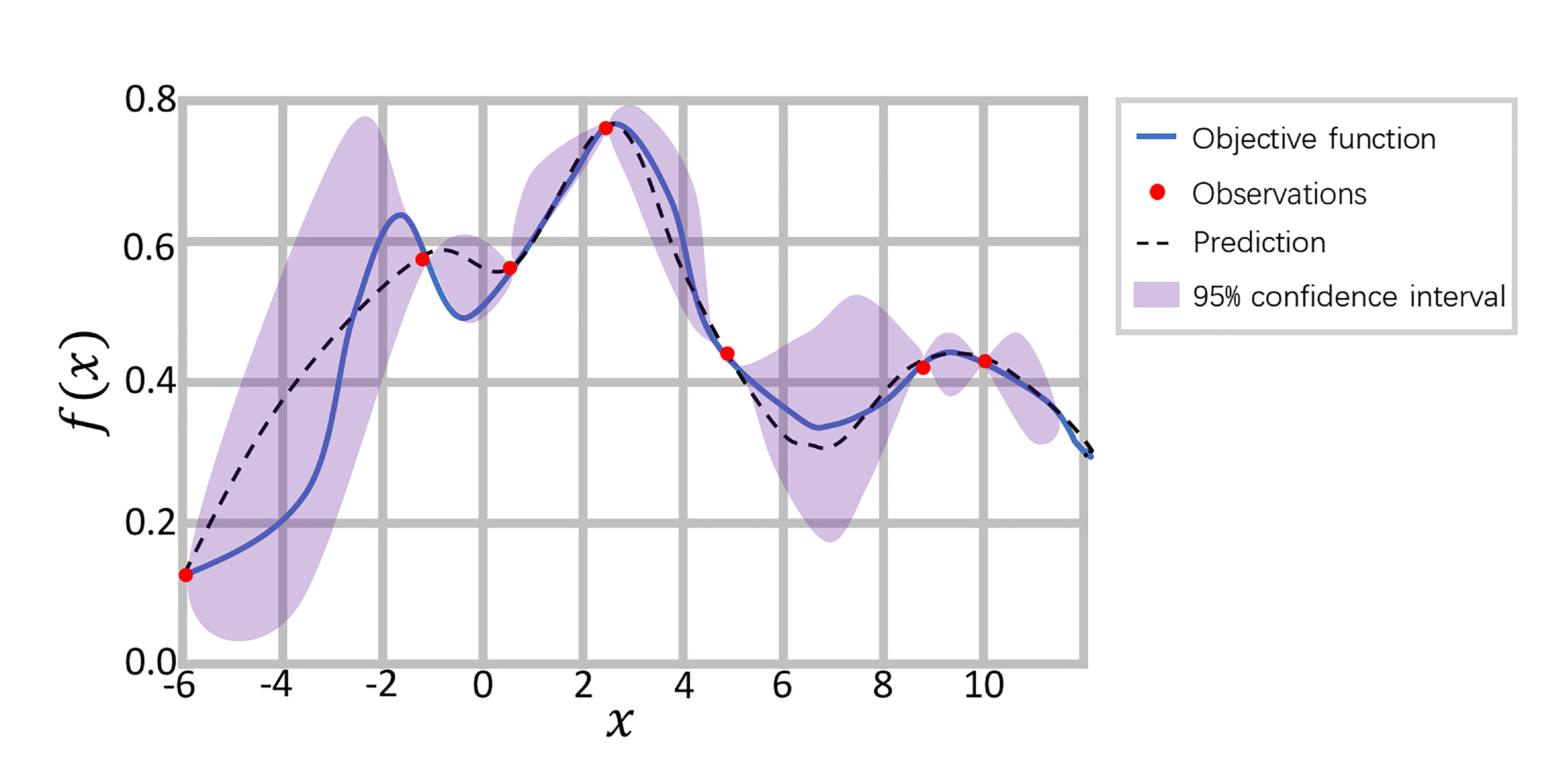}
\end{overpic}
\caption{Illustration of the Bayesian optimization procedure. Bayesian optimization estimates the true objective function with a probability model called a surrogate. The blue real curve represents the true objective function. Red cross points are sampling points for the true objective function $f(x)$. The black dotted curve and purple shadow represent the mean and confidence intervals estimated with the surrogate of the objective function.  }
\label{fig_BO}
\end{center}
\vspace{-0.5cm}
\end{figure}

\section{Bayesian optimization}
\label{section 3}
In this section, we present a brief introduction to Bayesian optimization.  Bayesian optimization is a powerful strategy for global optimization of objective functions that are expensive to evaluate~\cite{shahriari2015taking,bergstra2011algorithms}. This method is gaining great popularity in hyperparameter optimization. In particular, hyperparameter optimization in machine learning can be represented as follows:
\begin{equation}
	\begin{aligned}
	 x^{\star}=\arg \min _{x \in \mathcal{X}} f(x),	
	\end{aligned}
\end{equation}
where $f(x): \mathcal{X} \rightarrow \mathbb{R}$ is an objective function to minimize, $x^{\star}$ is a hyperparameter vector yielding the lowest value of $f$, and the dimension of domain $\mathcal{X}$ depends on the total type of concerned hyperparameters. In practice, the evaluation of the objective function is extremely costly, which leads to selecting proper hyperparameters by hand becoming intractable. Beyond the manual tuning method, grid search and random search~\cite{bergstra2012random} are two common methods that perform slightly better. However, these methods still waste a large amount of time evaluating poor hyperparameters across the entire search space, which is relatively inefficient. In contrast, Bayesian optimization estimates the true objective function with a probability model. Then, it utilizes Bayes' theorem to update this model based on previous results and chooses the next promising hyperparameters. In practice, this method can find better hyperparameters in less time. Figure~\ref{fig_BO} illustrates the Bayesian optimization procedure. 

%%%%%%%%%%%%%%%%%%
\begin{algorithm}[t]
\caption{Sequential Model-Based Optimization}\label{alalgorithm1}
\begin{algorithmic}[1]
\State $\mathcal{H}_0 \leftarrow \emptyset $
		\For {$n=1,2,\ldots,N$}
			\State $x_{n+1}=\arg \max_{x} \alpha_{n}(x,S_n)$
			\State evaluate $y_{n+1} = f(x_{n+1})$
			\State update $\mathcal{H}_{n+1} = \{\mathcal{H}_{n},(x_{n+1},y_{n+1})\}$
              \State fit a new model $S_n$ to $\mathcal{H}_{n+1}$
		\EndFor
\State \textbf{Return} $\mathcal{H}_{N}$
\end{algorithmic}
\end{algorithm}

Sequential model-based optimization (SMBO) algorithms are formalizations of Bayesian optimization.~\cite{bergstra2011algorithms} These algorithms have two key ingredients:

1. A probabilistic surrogate model $S$. SMBO approximates the objective function $f$ with a probabilistic model called a surrogate, which is cheaper to evaluate. This surrogate contains a prior distribution capturing beliefs about the behavior of the objective function and is then updated sequentially after each new trial.  

2. An acquisition function $\alpha: \mathcal{X} \rightarrow \mathbb{R}$. The acquisition function is the criterion by which the next vector of  hyperparameters is chosen from the surrogate function. 

For an SMBO algorithm at iteration n, the next location $x_{n+1}$ is selected by optimizing $\alpha_n$ and to evaluate the true $f$ to obtain a result $y_{n+1} = f(x_{n+1})$. The new tuple $(x_{n+1},y_{n+1})$ is appended to the historical set $\mathcal{H}$. Then, the surrogate model $S$ is updated incorporating the new results, which means that the prior is updated to produce a more informative posterior distribution over the space of objective functions. The pseudocode of this framework is summarized in Algorithm 1.

\begin{figure*}[htbp]
   \begin{center}
   \includegraphics[width=1.4\columnwidth]{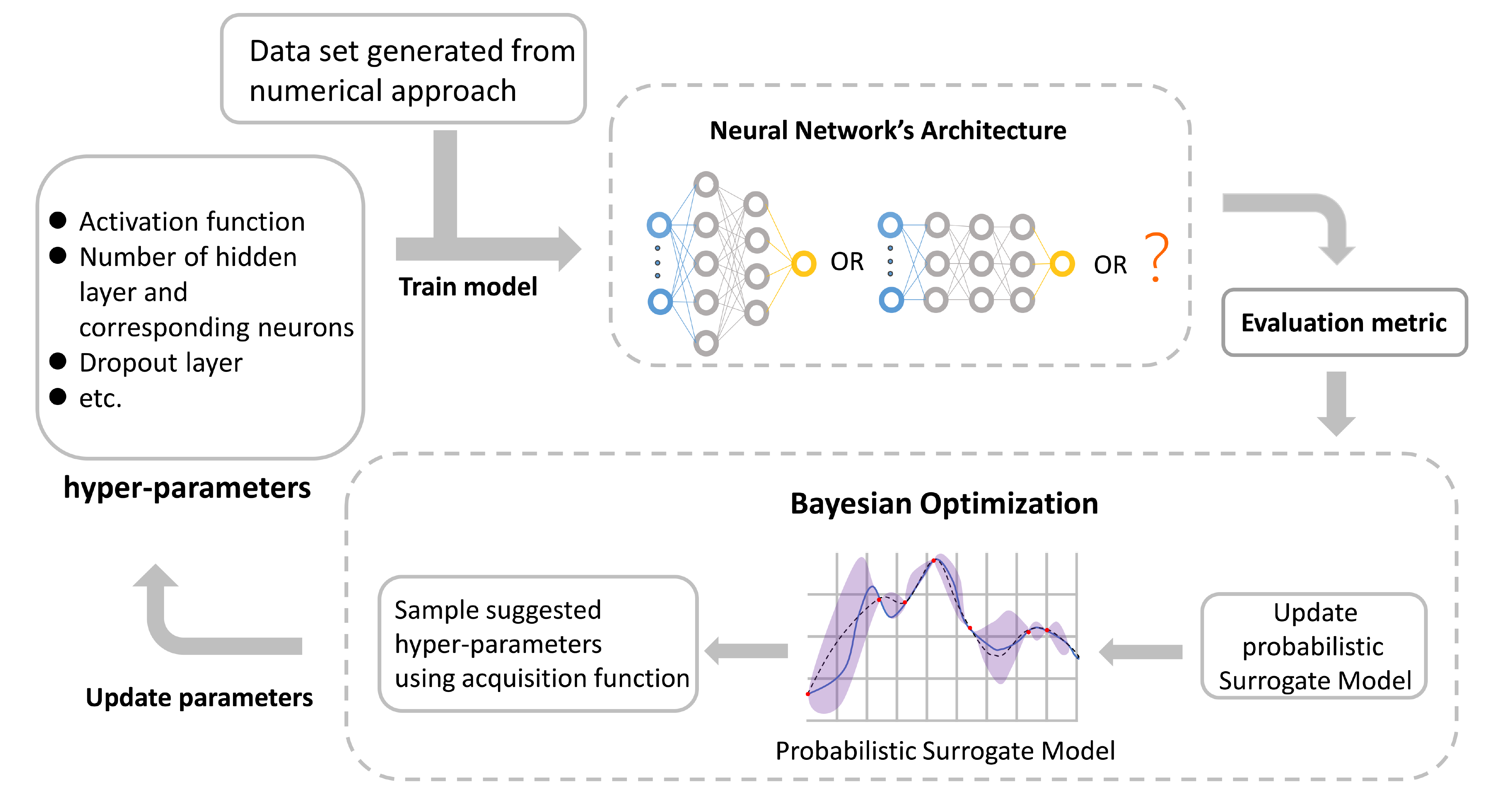}
		\caption{Schematic diagram of our neural network model combined with Bayesian optimization. The dataset training neural network to predict key rate is generated by some numerical approach. Here, the hyperparameters related to the neural network's architecture are not determined by humans but updated by Bayesian optimization. Bayesian optimization primarily establishes a probability model for the distribution of neural network hyperparameters and performance. Then, according to the evaluation metric produced in each trial, such as validation loss, Bayesian optimization updates the probabilistic surrogate model and suggests the next choice of hyperparameters. After several trials, we can automatically obtain the best-performing neural network.}
		\label{fig1}
     \end{center}
	\vspace{-0.2cm}
\end{figure*}

\begin{table*}[htbp]
     %\normalsize
     %\Huge
     \setlength{\abovecaptionskip}{-0.4cm}
	\begin{minipage}{\textwidth}
           \caption{Hyperparameter search space of the neural network under the TPE algorithm for the QPSK heterodyne detection protocol~\cite{lin2019asymptotic}. The neural network model we use here is a fully connected forward network. By fixing the number of neurons in the input layer and output layer, we search this neural network's architecture for hidden layers. For each of the three hidden layers, there are the number of neurons, activation function in this layer and the ratio of dropout layer following it waiting to determine, where the dropout technique~\cite{srivastava2014dropout} is used to prevent overfitting. The batch size of the training process and two essential hyperparameters $\gamma$ and $\varepsilon$ are searched as well. The brace $\{ \}$ refers to a finite set that contains all possible discrete values. Bracket $()$ represents a continuous range.} 
         
		\begin{center}
           \resizebox{\textwidth}{12mm}{
			\begin{tabular}{ccccccc}	
                \hline
                \hline
                	
				  &Number of neurons  & Activation function& Dropout&Batch size & $\gamma$&$\varepsilon$  \\
				\hline
				Input layer  & 29(fixed) & - & -&$\left\{32,64,128,256\right\}$ &$\left(0.05,0.2\right)$ &$\left(0.8,0.95\right)$\\
			     Hidden layer 1 & $\left\{512,1024\right\}$ &\{tanh,ReLU,sigmoid\} & $\left(0,0.3\right)$ & & &  \\
			     Hidden layer 2 & $\left\{128,256,512\right\}$ &\{tanh,ReLU,sigmoid\} & $\left(0,0.3\right)$ & & &\\
                     Hidden layer 3 & $\left\{128,256,512\right\}$ &\{tanh,ReLU,sigmoid\} & $\left(0,0.3\right)$ & & &\\
                     Output layer  & 1(fixed) &Linear(fixed) & -\\
             
                \hline
                \hline

			\end{tabular}}
		\end{center}
	\end{minipage}
	\vspace{-0.cm}
\end{table*} 

\begin{table*}
     %\Huge
     \normalsize
     \setlength{\abovecaptionskip}{-0.3cm}
	\begin{minipage}{\textwidth}
          \caption[Caption for LOF]{Hyperparameter search space of the neural network under the TPE algorithm for the QPSK homodyne detection protocol~\cite{PRXQuantum.2.040334}. Different from the QPSK heterodyne detection protocol~\cite{lin2019asymptotic}, here, we search the number of hidden layers in $3$ or $4$.}        
        
		\begin{center}
           \resizebox{\textwidth}{12mm}{
			\begin{tabular}{ccccccc}	
                \hline
                \hline
            
				  & Number of neurons & Activation function& Dropout&Batch size & $\gamma$&$\varepsilon$  \\
				\hline
				Input layer  & 29(fixed) & - & -&$\left[32,64,128,256\right]$ &$\left(0.05,0.2\right)$ &$\left(0.8,0.99\right)$\\
			     Hidden layer 1 & $\left\{128,256,512\right\}$ &[tanh,ReLU,sigmoid] & $\left(0,0.3\right)$ & & &  \\
			     Hidden layer 2 & $\left\{128,256,512\right\}$ &[tanh,ReLU,sigmoid] & $\left(0,0.3\right)$ & & &\\
                     Hidden layer 3 & $\left\{128,256\right\}$ &[tanh,ReLU,sigmoid] & $\left(0,0.3\right)$ & & &\\
                     Hidden layer 4(optional) & $\left\{64,128\right\}$ &[tanh,ReLU,sigmoid] & $\left(0,0.3\right)$ & & &\\
                     Output layer  & 1(fixed) &Linear(fixed) & -\\
            
                \hline
                \hline
           
			\end{tabular}}
		\end{center}
	\end{minipage}
	\vspace{-0.2cm}
\end{table*} 

The most common choice of acquisition function is expected improvement (EI): 
\begin{equation}
\mathrm{E} \mathrm{I}_{y^{*}}(x):=\int_{-\infty}^{\infty} \max \left(y^{*}-y, 0\right) p_{S}\left(y \mid x\right) d y
\end{equation}
Here $y^{*}$ is a threshold value of the objective function $f$, and $p_{S}\left(y\mid x\right)$ represents the surrogate probability model. If this expectation is positive, then the vector of hyprparameters $x$ is expected to produce a better result than $y^{*}$. There are several different strategies for constructing the surrogate model: a Gaussian process approach~\cite{williams2006gaussian}, random forests~\cite{breiman2001random} and a tree-structured Parzen estimator(TPE)~\cite{bergstra2011algorithms}. In this work, the TPE approach is adopted, which supports continuous, categorical and conditional parameters, as well as priors for each hyperparameter over which values are expected to perform best~\cite{hutter2015beyond}. In contrast, the Gaussian process approach and random forests only support one or two types of the above parameters, which are not capable of our following task covering continuous, categorical and conditional parameters. Instead of directly modeling $p\left(y\mid x\right)$, this method models $p\left(x\mid y\right)$ using two such densities over the configuration space $\mathcal{X}$:
\begin{equation}
p\left(x \mid y\right)= \begin{cases}\ell(x) & \text { if } y<y^{*} \\ g(x) & \text { if } y \geq y^{*}\end{cases}
\end{equation}
This algorithm chooses $y^*$ to be some quantile $\gamma$ of the observed y values, which means $p\left(y<y^*\right) = \gamma$. So the $E I_{y^{*}}(x)=\frac{\gamma y^{*} \ell(x)-\ell(x) \int_{-\infty}^{y^{*}} p(y) d y}{\gamma \ell(x)+(1-\gamma) g(x)} \propto\left(\gamma+\frac{g(x)}{\ell(x)}(1-\gamma)\right)^{-1}$. The tree-structured form of $\ell$ and $g$ makes it easy to draw many candidates according to $g(x) / \ell(x)$. On each iteration, the algorithm returns the candidate $x$ with the greatest EI. We implement this algorithm for the hyperparameter optimization of the neural networks predicting CV QKD key rates, by using a Python library called Hyperopt~\cite{bergstra2013hyperopt}.

\section{Method}
\label{section 4}

Artificial neural networks can approximate arbitrary bounded continuous mapping on a given domain, according to universal approximation theorem~\cite{hornik1989multilayer}. Therefore, we expect that the neural network can learn the mapping between input variables defined in the constraints of Eq.(3) and output key rates, which avoids solving the time-consuming optimization problem and computes key rates with low latency. We demonstrated this possibility of using a neural network to predict the key rates of discrete-modulated CV QKD in  previous work~\cite{zhou2021machine}. In that work, we built a four-layer fully connected  forward neural network holding a loss function designed specifically to predict the key rates of discrete-modulated CV QKD with homodyne detection. The objective loss function is the key ingredient to keep the output key rates reliable and tight. We retain it in this work but utilize the TPE algorithm to search other parts of the neural network to improve the network's overall performance. The specific formula of the loss function is as follows:

\begin{equation}
\begin{aligned}
\mathcal{L} &=\frac{1}{n} \sum_{i=1}^{n} \gamma\left(e_{i}^{* 2}+\max \left(e_{i}^{*},-\log _{10}(\varepsilon)\right)\right)\\
&-(1-\gamma)\left(\min \left(e_{i}^{*}, 0\right)\right)
\end{aligned}
\end{equation}

\begin{table*}
      %\Huge
      \normalsize
      \setlength{\abovecaptionskip}{-0.cm}
      \begin{minipage}{\textwidth}
	
           \caption{Resulting structure of neural networks of QPSK  heterodyne detection protocol~\cite{lin2019asymptotic}}

		\begin{center}
           \resizebox{\textwidth}{12mm}{
			\begin{tabular}{ccccccc}	
                \hline
                \hline
               	
				  & Number of neurons & Activation function& Dropout&Batch size & $\gamma$&$\varepsilon$  \\
				\hline
				Input layer  & 29& - & -&$64$ &$0.0539$ &$0.8727$\\
			     Hidden layer 1 & $1024$ &sigmoid & $0.0769$ & & &  \\
			     Hidden layer 2 & $256$ &tanh & $0.0362$ & & &\\
                     Hidden layer 3 & $256$ &sigmoid& $0.0481$ & & &\\
                     Output layer  & 1 &Linear & -\\
             
                \hline
                \hline

			\end{tabular}}
		\end{center}
	\end{minipage}
	\vspace{-0cm}
\end{table*} 

\begin{table*}
       %\Huge
       \normalsize
       \setlength{\abovecaptionskip}{-0.cm}
       \begin{minipage}{\textwidth}
	
           \caption{Resulting structure of neural networks of QPSK homodyne detection protocol~\cite{PRXQuantum.2.040334}} 

		\begin{center}
           \resizebox{\textwidth}{12mm}{
			\begin{tabular}{ccccccc}	
                \hline
                \hline
              	
				  & Number of neurons & Activation function& Dropout&Batch size & $\gamma$&$\varepsilon$  \\
				\hline
				Input layer  & 29& - & -&$64$ &$0.1227$ &$0.8784$\\
			     Hidden layer 1 & $512$ &tanh & $0.2210$ & & &  \\
			     Hidden layer 2 & $128$ &tanh & $0.2361$ & & &\\
                     Hidden layer 3 & $256$ &sigmoid& $0.0657$ & & &\\
                     Hidden layer 4 & $128$ &tanh& $0.0036$ & & &\\
                     Output layer  & 1 &Linear & -\\
              
                \hline
                \hline

			\end{tabular}}
		\end{center}
	\end{minipage}
	\vspace{-0.1cm}
\end{table*} 

\begin{figure*}
   \begin{center}
   \includegraphics[width=1.6\columnwidth]{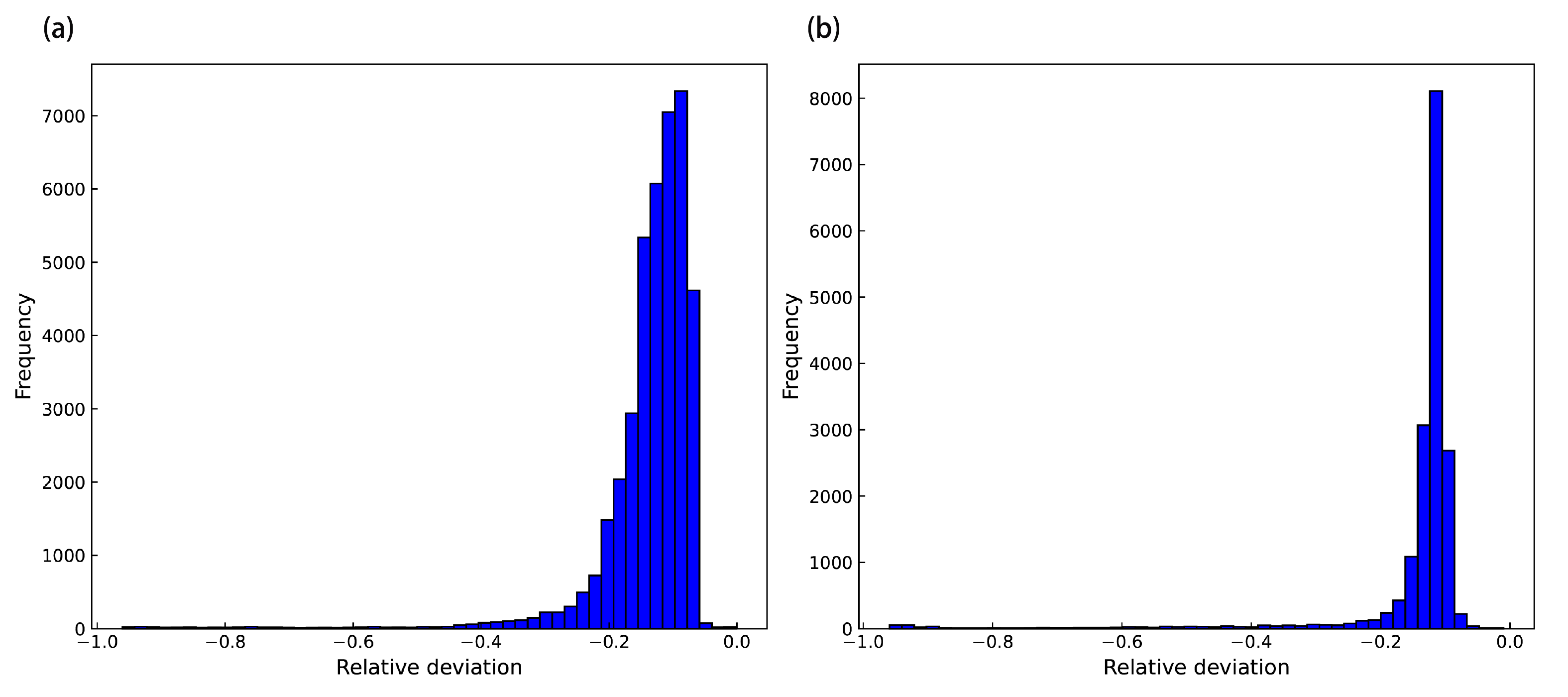}
\caption{Relative deviation distribution of key rates predicted securely in respective test set for two protocols: (a)Protocol 1(with heterogyne detection)~\cite{lin2019asymptotic}. In these secure results, the relative deviations falling in $\left[-20\%,0\right]$ account for $92.49\%$ and falling in $\left[-40\%,0\right]$ account for $96.48\%$. (b)Protocol 2 with impoved homodyne detection~\cite{PRXQuantum.2.040334}. The relative deviations falling in $\left[-20\%,0\right]$ account for $90.26\%$ and falling in $\left[-40\%,0\right]$ account for $98.39\%$.}
\label{fig2}
     \end{center}
	\vspace{-0.4cm}
\end{figure*}

\begin{figure*}
	\begin{center}
		\includegraphics[width=1.6\columnwidth]{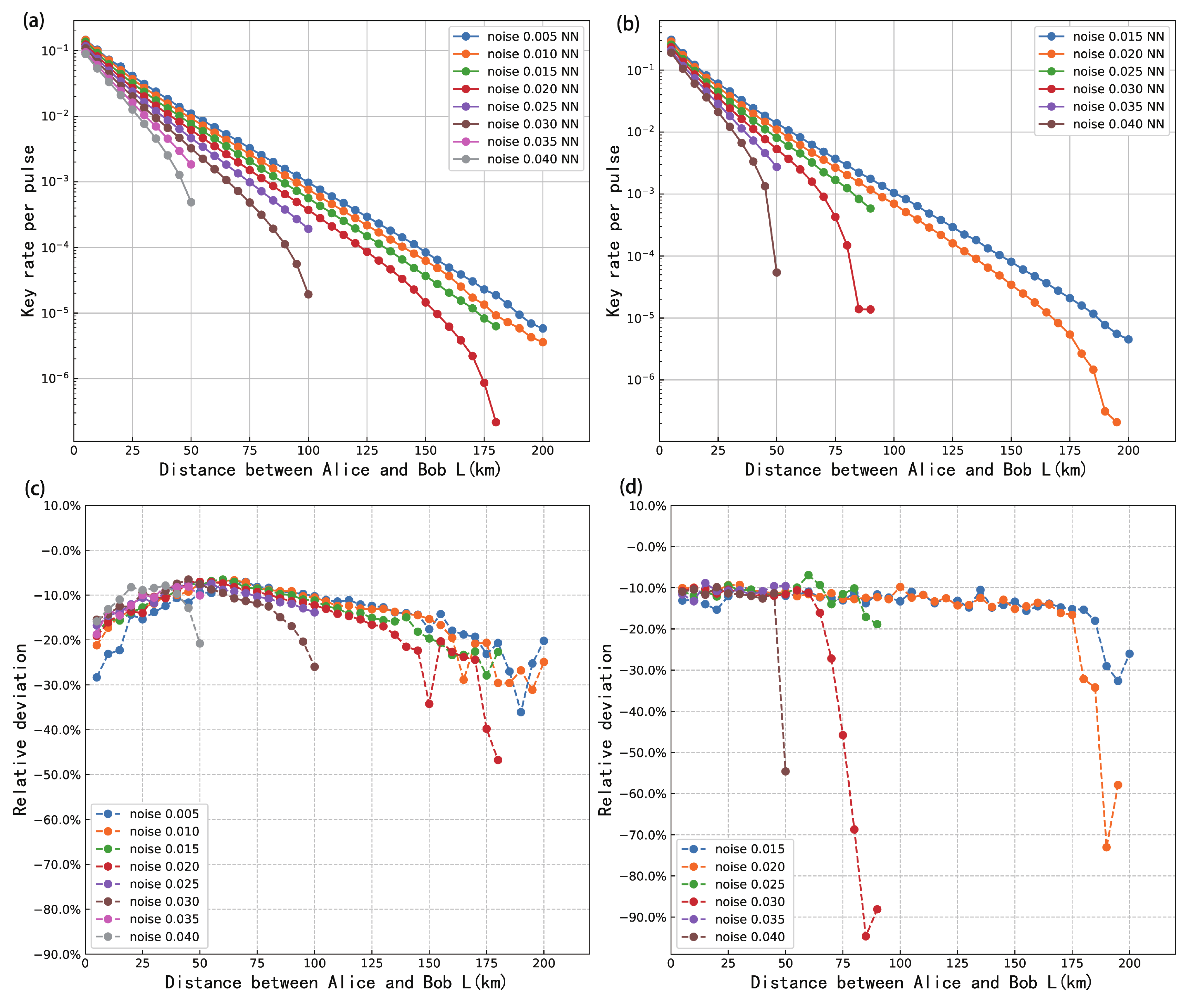}
		\caption{Comparison between NN-predicted results and numerical results in key rates for two protocols. (a)The predicted key rates vs. transmission distance for the QPSK heterodyne detection protocol~\cite{lin2019asymptotic} for different values of the excess noise, from top to bottom, $\xi=0.005,0.010,0.015,0.020,0.025,0.030,0.035,0.040$. The signal state amplitude of the QPSK heterodyne detection protocol is optimized in the range $\left[0.62,0.72\right]$ with a step of $0.01$. (b)The predicted key rates vs. transmission distance for the QPSK homodyne detection protocol~\cite{PRXQuantum.2.040334} for different values of excess noise, from top to bottom, $\xi=0.015,0.020,0.025,0.030,0.035,0.040$. The amplitude of the QPSK homodyne detection protocol is optimized in the range $\left[0.62,1.03\right]$ with a step of $0.01$. (c) and (d)Relative deviation between NN-predicted results and numerical results, respectively, for the QPSK heterodyne detection protocol and QPSK homodyne detection protocol, in which the corresponding NN-predicted results are shown in (a) and (b). For all points, we set reconciliation efficiency $\beta=0.95$, postselection $\Delta=0$, transmittance $\eta=10^{-\frac{a L}{10}}$ in the distance L with $a=0.2$ dB/km, photon-number cutoff $N_c=12$ and the maximal $N_i=300$ iteration of the first step in the numerical method. }
		\label{fig3}
	
       \end{center}
	
\end{figure*}

\begin{figure*}
	\begin{center}
	\includegraphics[width=1.8\columnwidth]{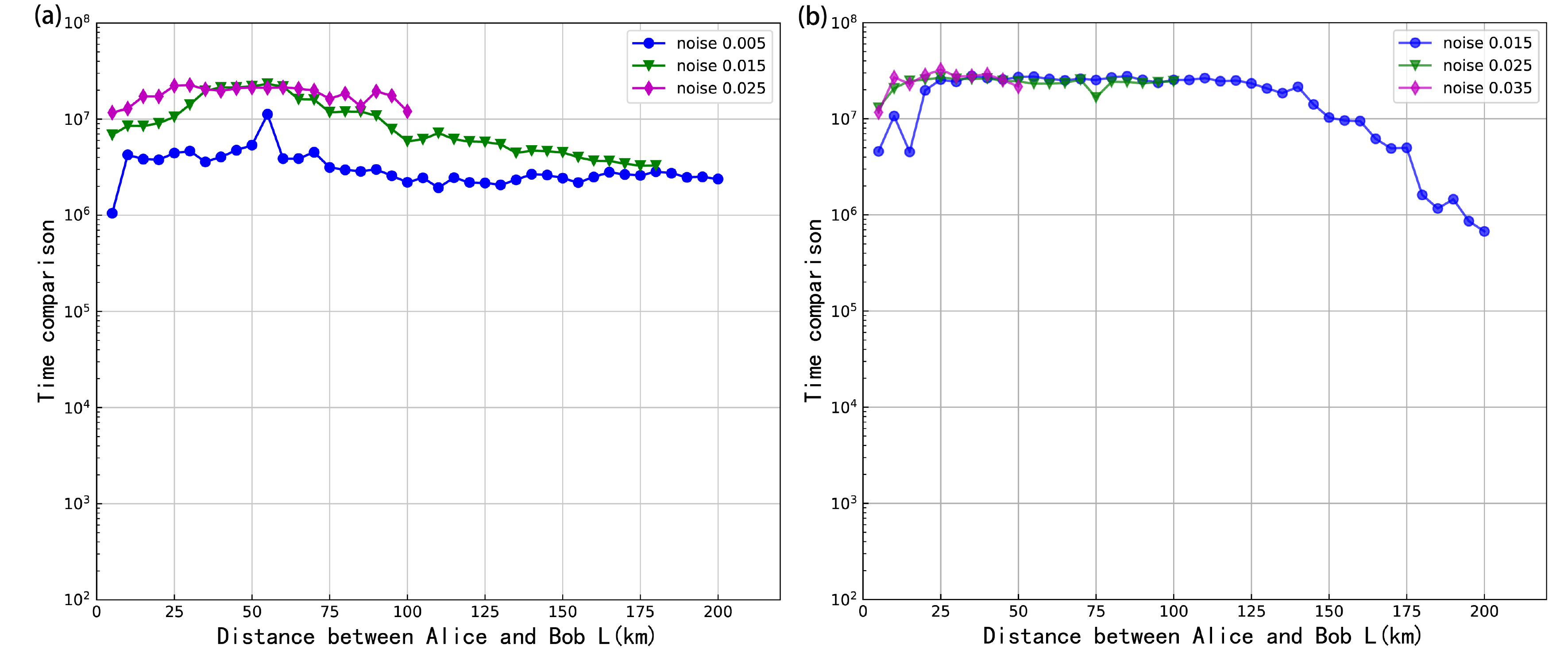}
		\caption{Running time comparison between neural-network method and numerical method. (a)Time comparison of protocol 1 for excess noise $\xi=0.005,0.015,0.025$ is shown as circles, triangles and diamonds respectively. (b)Time comparison of protocol 2 for excess noise $\xi=0.015,0.025,0.035$ is shown as circles, triangles and diamonds respectively. Each point refers to a ratio of time consumed by numerical method and neural-network method. The values of $\beta$, $\Delta$, $\eta$, $N_c$ and $N_i$ remain the same as those mentioned in Fig.~\ref{fig3}. }
		\label{fig4}
	\end{center}
	\vspace{-1cm}
\end{figure*}

For training inputs $\left\{\vec{x}_{i}\right\}$ and corresponding labels $\left\{{y}_{i}\right\}$, here $n$ is the size of $\left\{\vec{x}_{i}\right\}$, $e_i^*=y_{i}^{*p}-y_{i}^{ *}$ is the residual error between preprocessed label $y_{i}^{*}$ and the corresponding output of the neural network $y_{i}^{* p}$, where  $y_{i}^{ *}=-\log _{10}\left(y_{i}\right)$. There are two significant hyperparameters $\gamma$ and $\varepsilon$ contained in this loss function, the choices of which are crucial to a model's performance, as we presented in Ref.\cite{zhou2021machine}. The meaning of hyperparameter $\gamma$ is to force the predicted key rate as information-theoretically secure as possible, and $\varepsilon$ is to force the predicted key rate as close to numerical results as possible. Here, apart from the input layer and output layer, we do not fix the structure of the neural network but utilize the TPE algorithm to search it efficiently in a set configuration space. An illustration of our model is shown in Fig.~\ref{fig1}.

\section{Result}
\label{section 5}

After this training under TPE searching is complete, we obtain the resulting structures of neural networks in both cases, which are shown in Table 3-4. Then, we use the selected and trained network to predict key rates on the test set for both protocols. The predicted key rates that show security achieve as high as $99.15\%$ for the QPSK  heterodyne detection protocol and $99.59\%$ for the QPSK homodyne detection protocol, which suggests that our method combining a neural network with Bayesian optimization is highly reliable. For those key rates predicted securely, namely, predicted results are lower than the true values, we plot their relative deviation distributions for both protocols in Fig.~\ref{fig2}. Figure~\ref{fig2} suggests that our method has good tightness.

Before training the neural network under the TPE method, we generate datasets for two different protocols by the aforementioned numerical approach. To obtain datasets with diversity, for the QPSK  heterodyne detection protocol,  we generate $36$ sets of data from excess noise $\xi=0.0045-0.0405$. Each dataset contains $80$ random samplings for $\xi$ from an interval of length $0.001$, for example $\left[0.0045,0.0055\right]$. Under each random sampling, we generate data every $5$km with the transmission distance $L$ up to $200 $km. At each distance, we generate data from amplitude $\alpha=0.62-0.72$ in a  step of $0.01$. The total datasets contain $809,600$ input instances $\left\{\vec{x}_{i}\right\}$ and corresponding labels $\left\{{y}_{i}\right\}$. For the QPSK homodyne detection protocol, excess noise is sampled randomly from $\xi = 0.014-0.042$, where the length of sampling interval is $0.002$, for example $\left[0.014,0.016\right]$, and amplitude $\alpha$ is sampled from $\left[0.60,1.05\right]$. The size of total datasets is $368,116$. For both protocols, each  $\vec{x}_{i} \in \left\{\vec{x}_{i}\right\}$ represents a vector of $29$ variables, there are $16$ variables that are the right parts of the first four constraints of Eq.~\ref{op}, $12$ variables are nondiagonal elements of the right side matrix of the last constraint of Eq.~\ref{op}, and the remaining variable is excess noise $\xi$. Label $y_i \in \left\{\vec{y}_{i}\right\}$ represents the corresponding key rate. Therefore, we fix the neurons of the network's input layer in $29$ and output layer in $1$, and the search space of other hyperparameters can be found in Table 1-2.

Before feeding data into neural networks, we split data into a training set and a test set and implement data preprocessing as in Ref.\cite{zhou2021machine}. For the QPSK heterodyne detection protocol, the training set contains $769,120$ data instances, and the test set contains $40,480$ data instances. For the QPSK homodyne detection protocol, the training set contains $327,636$ data instances, and the test set contains $17,244$ data instances. 
For both cases, there is $10\%$ of the training data split as the validation set. We generate the dataset on the blade cluster system of the High Performance Computing Center of Nanjing University. We consume over $250,000$ core hours, and the node we use contains 4 Intel Xeon Gold 6248 CPUs, which involves immense computational power. Under the TPE algorithm with max iteration $10$, the Adam algorithm~\cite{kingma2014adam} is used to train neural networks for $200$ epochs, and the initial learning rate is set to $0.001$. It takes roughly $53$ hours for the QPSK  heterodyne detection protocol and $23$ hours for the QPSK homodyne detection protocol on an Nvidia A100 GPU.

Here, we also compare the predicted results with numerical results in key rates versus transmission distance for two protocols. The comparison is shown in Fig.~\ref{fig3}. For this plot, we implement the same numerical approach to compute the best key rates of two protocols for different excess noises by optimizing the amplitude $\alpha$ of signal states in the range $\left[0.62,0.72\right]$ and $\left[0.62,1.03\right]$ with a step of $0.01$. The choice of the excess noise range is consistent with the sampling interval of previous training data. The photon-number cutoff $N_c$ is $12$, and the maximal iteration number of the first step in the numerical approach $N_i=300$. We record the corresponding $29$ variables producing the best key rates as neural networks' inputs to predict key rates. As shown in Figs.~\ref{fig3}(c) and (d), the predicted results are all secure and remain tight with relative deviations between $10\%$ and $20\%$ when the transmission distance is below $150$ km for both protocols.

To show the efficiency of our method, we compare the running time between the neural network method and the numerical method on a high-performance personal computer with a 3.3 GHz AMD Ryzen 9 4900H 16 GB of RAM, as shown in Fig.~\ref{fig4}. The results suggest that the neural network method is generally 6-8 orders of magnitude of the numerical methods. For example, when $\xi=0.025$, the numerical method consumes approximately $850$ seconds to calculate the key rate at $50$ km for the QPSK heterodyne detection protocol. When $\xi=0.035$, the numerical method consumes approximately $1260$ seconds at $25$ km to calculate the key rate for the QPSK homodyne detection protocol. However, we can use a trained neural network to obtain results in approximately $0.0001$ seconds, which is almost real time.

\section{Discussion and conclusion}
\label{section 6}
To summarize, we develop a neural network model combined with Bayesian optimization to directly extract key rates with high reliability, considerable tightness and great efficiency. Beyond designing the neural network architecture by human and troublesome manual tuning of hyperparameters, we utilize a special Bayesian optimization method called the TPE algorithm to automatically search the structure and hyperparameters that are the best fit for a given dataset. We exemplify our method on two promising discrete-modulated CV QKD protocols varied by different detection techniques across a large range of excess noise and transmission distances. For both protocols, the neural networks selected by the TPE algorithm predict the information-theoretically secure key rates with great high probability(up to $99.15\%$ for the QPSK heterodyne detection protocol and $99.59\%$ for the QPSK homodyne detection protocol), and the results present considerable tightness.

We show that our method can achieve approximately $10^7$ faster than the numerical method, which completely satisfies the requirement of the QKD system in practice. In contrast, the numerical method takes several minutes to calculate a point of key rate, which is intolerable since many free-space sessions, such as satellite-ground or handheld QKD might have a window of only minutes. While collecting enough data based on the numerical method to train the model consumes a large amount of computing power, we can consider these large computations offline. Once we obtain the trained neural network, it can be deployed on a certain device to infer key rates online in milliseconds by giving new inputs from the experiment. Ref.\cite{wang2019machine} demonstrated that a neural network method for parameter optimization of QKD can be deployed on various mobile low-power systems, which brings advantages of more power efficiency and low latency. We can also forecast that our neural network method combined with Bayesian optimization will play an essential role in free-space QKD scenarios such as handheld~\cite{melen2017handheld}, drone-based~\cite{hill2017drone} or satellite-ground QKD~\cite{liao2017satellite}. Several works have focused on machine learning for optimal parameters in QKD~\cite{wang2019machine,liu2018integrating,lu2019parameter,ding2020predicting}. However, our work predicts secure key rates directly by automatically designed neural networks, which goes further than our previous work~\cite{zhou2021machine}. 

Based on our model, there are several directions worthy of investigation for future work. Up to now, we have only covered computing the asymptotic key rates. However, finite-size effects are practical issues considered in discrete-modulated CV-QKD~\cite{leverrier2010finite}. Note that a recent work has analyzed the security and performance of discrete-modulated CV-QKD under a finite-size scenario~\cite{almeida2021secret}, which inspires us to improve our model. To address these issues, we also consider applying our model to other protocols in future work. Moreover, the issue of post-processing (notably the error correction part) still limits the overall time acceleration for a discrete-modulated continuous-variable QKD system. Note that the error correction involving binary or quaternary error-correcting codes is less complex compared with the situation of Gaussian modulation. Therefore, we also consider developing an effective error-correction protocol for CV QKD with discrete modulation using machine learning techniques in the future.

\section*{Acknowledgments}
We gratefully acknowledge the support from the Natural Science Foundation of Jiangsu Province (No. BK20211145), the Fundamental Research Funds for the Central Universities (No. 020414380182), the Key Research and Development Program of Nanjing Jiangbei New Aera (No. ZDYD20210101), the Program for Innovative Talents and Entrepreneurs in Jiangsu(No. JSSCRC2021484), and the Key-Area Research and Development Program of Guangdong Province (No. 2020B0303040001). The authors would like to thank the High Performance Computing Center of Nanjing University for the numerical calculations.

%\bibliography{Ref}

\begin{thebibliography}{62}
\expandafter\ifx\csname natexlab\endcsname\relax\def\natexlab#1{#1}\fi
\expandafter\ifx\csname bibnamefont\endcsname\relax
  \def\bibnamefont#1{#1}\fi
\expandafter\ifx\csname bibfnamefont\endcsname\relax
  \def\bibfnamefont#1{#1}\fi
\expandafter\ifx\csname citenamefont\endcsname\relax
  \def\citenamefont#1{#1}\fi
\expandafter\ifx\csname url\endcsname\relax
  \def\url#1{\texttt{#1}}\fi
\expandafter\ifx\csname urlprefix\endcsname\relax\def\urlprefix{URL }\fi
\providecommand{\bibinfo}[2]{#2}
\providecommand{\eprint}[2][]{\url{#2}}

\bibitem[{\citenamefont{Grigorescu et~al.}(2020)\citenamefont{Grigorescu,
  Trasnea, Cocias, and Macesanu}}]{grigorescu2020survey}
\bibinfo{author}{\bibfnamefont{S.}~\bibnamefont{Grigorescu}},
  \bibinfo{author}{\bibfnamefont{B.}~\bibnamefont{Trasnea}},
  \bibinfo{author}{\bibfnamefont{T.}~\bibnamefont{Cocias}}, \bibnamefont{and}
  \bibinfo{author}{\bibfnamefont{G.}~\bibnamefont{Macesanu}},
  \bibinfo{journal}{Journal of Field Robotics} \textbf{\bibinfo{volume}{37}},
  \bibinfo{pages}{362} (\bibinfo{year}{2020}).

\bibitem[{\citenamefont{Levinson et~al.}(2011)\citenamefont{Levinson, Askeland,
  Becker, Dolson, Held, Kammel, Kolter, Langer, Pink, Pratt
  et~al.}}]{levinson2011towards}
\bibinfo{author}{\bibfnamefont{J.}~\bibnamefont{Levinson}},
  \bibinfo{author}{\bibfnamefont{J.}~\bibnamefont{Askeland}},
  \bibinfo{author}{\bibfnamefont{J.}~\bibnamefont{Becker}},
  \bibinfo{author}{\bibfnamefont{J.}~\bibnamefont{Dolson}},
  \bibinfo{author}{\bibfnamefont{D.}~\bibnamefont{Held}},
  \bibinfo{author}{\bibfnamefont{S.}~\bibnamefont{Kammel}},
  \bibinfo{author}{\bibfnamefont{J.~Z.} \bibnamefont{Kolter}},
  \bibinfo{author}{\bibfnamefont{D.}~\bibnamefont{Langer}},
  \bibinfo{author}{\bibfnamefont{O.}~\bibnamefont{Pink}},
  \bibinfo{author}{\bibfnamefont{V.}~\bibnamefont{Pratt}},
  \bibnamefont{et~al.}, in \emph{\bibinfo{booktitle}{2011 IEEE intelligent
  vehicles symposium (IV)}} (\bibinfo{organization}{IEEE},
  \bibinfo{year}{2011}), pp. \bibinfo{pages}{163--168}.

\bibitem[{\citenamefont{Deng et~al.}(2013)\citenamefont{Deng, Hinton, and
  Kingsbury}}]{deng2013new}
\bibinfo{author}{\bibfnamefont{L.}~\bibnamefont{Deng}},
  \bibinfo{author}{\bibfnamefont{G.}~\bibnamefont{Hinton}}, \bibnamefont{and}
  \bibinfo{author}{\bibfnamefont{B.}~\bibnamefont{Kingsbury}}, in
  \emph{\bibinfo{booktitle}{2013 IEEE international conference on acoustics,
  speech and signal processing}} (\bibinfo{organization}{IEEE},
  \bibinfo{year}{2013}), pp. \bibinfo{pages}{8599--8603}.

\bibitem[{\citenamefont{Young et~al.}(2018)\citenamefont{Young, Hazarika,
  Poria, and Cambria}}]{young2018recent}
\bibinfo{author}{\bibfnamefont{T.}~\bibnamefont{Young}},
  \bibinfo{author}{\bibfnamefont{D.}~\bibnamefont{Hazarika}},
  \bibinfo{author}{\bibfnamefont{S.}~\bibnamefont{Poria}}, \bibnamefont{and}
  \bibinfo{author}{\bibfnamefont{E.}~\bibnamefont{Cambria}},
  \bibinfo{journal}{ieee Computational intelligenCe magazine}
  \textbf{\bibinfo{volume}{13}}, \bibinfo{pages}{55} (\bibinfo{year}{2018}).

\bibitem[{\citenamefont{Jumper et~al.}(2021)\citenamefont{Jumper, Evans,
  Pritzel, Green, Figurnov, Ronneberger, Tunyasuvunakool, Bates,
  {\v{Z}}{\'\i}dek, Potapenko et~al.}}]{jumper2021highly}
\bibinfo{author}{\bibfnamefont{J.}~\bibnamefont{Jumper}},
  \bibinfo{author}{\bibfnamefont{R.}~\bibnamefont{Evans}},
  \bibinfo{author}{\bibfnamefont{A.}~\bibnamefont{Pritzel}},
  \bibinfo{author}{\bibfnamefont{T.}~\bibnamefont{Green}},
  \bibinfo{author}{\bibfnamefont{M.}~\bibnamefont{Figurnov}},
  \bibinfo{author}{\bibfnamefont{O.}~\bibnamefont{Ronneberger}},
  \bibinfo{author}{\bibfnamefont{K.}~\bibnamefont{Tunyasuvunakool}},
  \bibinfo{author}{\bibfnamefont{R.}~\bibnamefont{Bates}},
  \bibinfo{author}{\bibfnamefont{A.}~\bibnamefont{{\v{Z}}{\'\i}dek}},
  \bibinfo{author}{\bibfnamefont{A.}~\bibnamefont{Potapenko}},
  \bibnamefont{et~al.}, \bibinfo{journal}{Nature}
  \textbf{\bibinfo{volume}{596}}, \bibinfo{pages}{583} (\bibinfo{year}{2021}).

\bibitem[{\citenamefont{Davies et~al.}(2021)\citenamefont{Davies,
  Veli{\v{c}}kovi{\'c}, Buesing, Blackwell, Zheng, Toma{\v{s}}ev, Tanburn,
  Battaglia, Blundell, Juh{\'a}sz et~al.}}]{davies2021advancing}
\bibinfo{author}{\bibfnamefont{A.}~\bibnamefont{Davies}},
  \bibinfo{author}{\bibfnamefont{P.}~\bibnamefont{Veli{\v{c}}kovi{\'c}}},
  \bibinfo{author}{\bibfnamefont{L.}~\bibnamefont{Buesing}},
  \bibinfo{author}{\bibfnamefont{S.}~\bibnamefont{Blackwell}},
  \bibinfo{author}{\bibfnamefont{D.}~\bibnamefont{Zheng}},
  \bibinfo{author}{\bibfnamefont{N.}~\bibnamefont{Toma{\v{s}}ev}},
  \bibinfo{author}{\bibfnamefont{R.}~\bibnamefont{Tanburn}},
  \bibinfo{author}{\bibfnamefont{P.}~\bibnamefont{Battaglia}},
  \bibinfo{author}{\bibfnamefont{C.}~\bibnamefont{Blundell}},
  \bibinfo{author}{\bibfnamefont{A.}~\bibnamefont{Juh{\'a}sz}},
  \bibnamefont{et~al.}, \bibinfo{journal}{Nature}
  \textbf{\bibinfo{volume}{600}}, \bibinfo{pages}{70} (\bibinfo{year}{2021}).

\bibitem[{\citenamefont{Gao et~al.}(2018)\citenamefont{Gao, Qiao, Jiao, Ma, Hu,
  Ren, Yang, Tang, Yung, and Jin}}]{gao2018experimental}
\bibinfo{author}{\bibfnamefont{J.}~\bibnamefont{Gao}},
  \bibinfo{author}{\bibfnamefont{L.-F.} \bibnamefont{Qiao}},
  \bibinfo{author}{\bibfnamefont{Z.-Q.} \bibnamefont{Jiao}},
  \bibinfo{author}{\bibfnamefont{Y.-C.} \bibnamefont{Ma}},
  \bibinfo{author}{\bibfnamefont{C.-Q.} \bibnamefont{Hu}},
  \bibinfo{author}{\bibfnamefont{R.-J.} \bibnamefont{Ren}},
  \bibinfo{author}{\bibfnamefont{A.-L.} \bibnamefont{Yang}},
  \bibinfo{author}{\bibfnamefont{H.}~\bibnamefont{Tang}},
  \bibinfo{author}{\bibfnamefont{M.-H.} \bibnamefont{Yung}}, \bibnamefont{and}
  \bibinfo{author}{\bibfnamefont{X.-M.} \bibnamefont{Jin}},
  \bibinfo{journal}{Phys. Rev. Lett.} \textbf{\bibinfo{volume}{120}},
  \bibinfo{pages}{240501} (\bibinfo{year}{2018}).

\bibitem[{\citenamefont{Yang et~al.}(2019)\citenamefont{Yang, Ren, Ma, Xiao,
  Ye, Song, Xu, Yung, Li, and Guo}}]{yang2019experimental}
\bibinfo{author}{\bibfnamefont{M.}~\bibnamefont{Yang}},
  \bibinfo{author}{\bibfnamefont{C.-l.} \bibnamefont{Ren}},
  \bibinfo{author}{\bibfnamefont{Y.-c.} \bibnamefont{Ma}},
  \bibinfo{author}{\bibfnamefont{Y.}~\bibnamefont{Xiao}},
  \bibinfo{author}{\bibfnamefont{X.-J.} \bibnamefont{Ye}},
  \bibinfo{author}{\bibfnamefont{L.-L.} \bibnamefont{Song}},
  \bibinfo{author}{\bibfnamefont{J.-S.} \bibnamefont{Xu}},
  \bibinfo{author}{\bibfnamefont{M.-H.} \bibnamefont{Yung}},
  \bibinfo{author}{\bibfnamefont{C.-F.} \bibnamefont{Li}}, \bibnamefont{and}
  \bibinfo{author}{\bibfnamefont{G.-C.} \bibnamefont{Guo}},
  \bibinfo{journal}{Phys. Rev. Lett.} \textbf{\bibinfo{volume}{123}},
  \bibinfo{pages}{190401} (\bibinfo{year}{2019}).

\bibitem[{\citenamefont{Ahmed et~al.}(2021)\citenamefont{Ahmed, Mu{\~n}oz,
  Nori, and Kockum}}]{ahmed2021classification}
\bibinfo{author}{\bibfnamefont{S.}~\bibnamefont{Ahmed}},
  \bibinfo{author}{\bibfnamefont{C.~S.} \bibnamefont{Mu{\~n}oz}},
  \bibinfo{author}{\bibfnamefont{F.}~\bibnamefont{Nori}}, \bibnamefont{and}
  \bibinfo{author}{\bibfnamefont{A.~F.} \bibnamefont{Kockum}},
  \bibinfo{journal}{Phys. Rev. Research} \textbf{\bibinfo{volume}{3}},
  \bibinfo{pages}{033278} (\bibinfo{year}{2021}).

\bibitem[{\citenamefont{Bukov et~al.}(2018)\citenamefont{Bukov, Day, Sels,
  Weinberg, Polkovnikov, and Mehta}}]{bukov2018reinforcement}
\bibinfo{author}{\bibfnamefont{M.}~\bibnamefont{Bukov}},
  \bibinfo{author}{\bibfnamefont{A.~G.} \bibnamefont{Day}},
  \bibinfo{author}{\bibfnamefont{D.}~\bibnamefont{Sels}},
  \bibinfo{author}{\bibfnamefont{P.}~\bibnamefont{Weinberg}},
  \bibinfo{author}{\bibfnamefont{A.}~\bibnamefont{Polkovnikov}},
  \bibnamefont{and} \bibinfo{author}{\bibfnamefont{P.}~\bibnamefont{Mehta}},
  \bibinfo{journal}{Phys. Rev. X} \textbf{\bibinfo{volume}{8}},
  \bibinfo{pages}{031086} (\bibinfo{year}{2018}).

\bibitem[{\citenamefont{Lumino et~al.}(2018)\citenamefont{Lumino, Polino, Rab,
  Milani, Spagnolo, Wiebe, and Sciarrino}}]{lumino2018experimental}
\bibinfo{author}{\bibfnamefont{A.}~\bibnamefont{Lumino}},
  \bibinfo{author}{\bibfnamefont{E.}~\bibnamefont{Polino}},
  \bibinfo{author}{\bibfnamefont{A.~S.} \bibnamefont{Rab}},
  \bibinfo{author}{\bibfnamefont{G.}~\bibnamefont{Milani}},
  \bibinfo{author}{\bibfnamefont{N.}~\bibnamefont{Spagnolo}},
  \bibinfo{author}{\bibfnamefont{N.}~\bibnamefont{Wiebe}}, \bibnamefont{and}
  \bibinfo{author}{\bibfnamefont{F.}~\bibnamefont{Sciarrino}},
  \bibinfo{journal}{Phys. Rev. Appl.} \textbf{\bibinfo{volume}{10}},
  \bibinfo{pages}{044033} (\bibinfo{year}{2018}).

\bibitem[{\citenamefont{Niu et~al.}(2019)\citenamefont{Niu, Boixo, Smelyanskiy,
  and Neven}}]{niu2019universal}
\bibinfo{author}{\bibfnamefont{M.~Y.} \bibnamefont{Niu}},
  \bibinfo{author}{\bibfnamefont{S.}~\bibnamefont{Boixo}},
  \bibinfo{author}{\bibfnamefont{V.~N.} \bibnamefont{Smelyanskiy}},
  \bibnamefont{and} \bibinfo{author}{\bibfnamefont{H.}~\bibnamefont{Neven}},
  \bibinfo{journal}{npj Quantum Inf.} \textbf{\bibinfo{volume}{5}},
  \bibinfo{pages}{33} (\bibinfo{year}{2019}).

\bibitem[{\citenamefont{Hentschel and Sanders}(2011)}]{hentschel2011efficient}
\bibinfo{author}{\bibfnamefont{A.}~\bibnamefont{Hentschel}} \bibnamefont{and}
  \bibinfo{author}{\bibfnamefont{B.~C.} \bibnamefont{Sanders}},
  \bibinfo{journal}{Phys. Rev. Lett.} \textbf{\bibinfo{volume}{107}},
  \bibinfo{pages}{233601} (\bibinfo{year}{2011}).

\bibitem[{\citenamefont{Bennett and Brassard}(1984)}]{bennett1984Quantum}
\bibinfo{author}{\bibfnamefont{C.~H.} \bibnamefont{Bennett}} \bibnamefont{and}
  \bibinfo{author}{\bibfnamefont{G.}~\bibnamefont{Brassard}}, in
  \emph{\bibinfo{booktitle}{Conf. on Computers, Systems and Signal Processing
  (Bangalore, India}} (\bibinfo{year}{1984}), vol. \bibinfo{volume}{175}.

\bibitem[{\citenamefont{Ekert}(1991)}]{ekert1991quantum}
\bibinfo{author}{\bibfnamefont{A.~K.} \bibnamefont{Ekert}},
  \bibinfo{journal}{Phys. Rev. Lett.} \textbf{\bibinfo{volume}{67}},
  \bibinfo{pages}{661} (\bibinfo{year}{1991}).

\bibitem[{\citenamefont{Shor and Preskill}(2000)}]{shor2000simple}
\bibinfo{author}{\bibfnamefont{P.~W.} \bibnamefont{Shor}} \bibnamefont{and}
  \bibinfo{author}{\bibfnamefont{J.}~\bibnamefont{Preskill}},
  \bibinfo{journal}{Phys. Rev. Lett.} \textbf{\bibinfo{volume}{85}},
  \bibinfo{pages}{441} (\bibinfo{year}{2000}).

\bibitem[{\citenamefont{Lo et~al.}(2014)\citenamefont{Lo, Curty, and
  Tamaki}}]{lo2014secure}
\bibinfo{author}{\bibfnamefont{H.-K.} \bibnamefont{Lo}},
  \bibinfo{author}{\bibfnamefont{M.}~\bibnamefont{Curty}}, \bibnamefont{and}
  \bibinfo{author}{\bibfnamefont{K.}~\bibnamefont{Tamaki}},
  \bibinfo{journal}{Nat. Photonics} \textbf{\bibinfo{volume}{8}},
  \bibinfo{pages}{595} (\bibinfo{year}{2014}).

\bibitem[{\citenamefont{Gisin et~al.}(2002)\citenamefont{Gisin, Ribordy,
  Tittel, and Zbinden}}]{gisin2002quantum}
\bibinfo{author}{\bibfnamefont{N.}~\bibnamefont{Gisin}},
  \bibinfo{author}{\bibfnamefont{G.}~\bibnamefont{Ribordy}},
  \bibinfo{author}{\bibfnamefont{W.}~\bibnamefont{Tittel}}, \bibnamefont{and}
  \bibinfo{author}{\bibfnamefont{H.}~\bibnamefont{Zbinden}},
  \bibinfo{journal}{Rev. Mod. Phys.} \textbf{\bibinfo{volume}{74}},
  \bibinfo{pages}{145} (\bibinfo{year}{2002}).

\bibitem[{\citenamefont{Grosshans and
  Grangier}(2002)}]{grosshans2002continuous}
\bibinfo{author}{\bibfnamefont{F.}~\bibnamefont{Grosshans}} \bibnamefont{and}
  \bibinfo{author}{\bibfnamefont{P.}~\bibnamefont{Grangier}},
  \bibinfo{journal}{Phys. Rev. Lett.} \textbf{\bibinfo{volume}{88}},
  \bibinfo{pages}{057902} (\bibinfo{year}{2002}).

\bibitem[{\citenamefont{Lance et~al.}(2005)\citenamefont{Lance, Symul, Sharma,
  Weedbrook, Ralph, and Lam}}]{lance2005no}
\bibinfo{author}{\bibfnamefont{A.~M.} \bibnamefont{Lance}},
  \bibinfo{author}{\bibfnamefont{T.}~\bibnamefont{Symul}},
  \bibinfo{author}{\bibfnamefont{V.}~\bibnamefont{Sharma}},
  \bibinfo{author}{\bibfnamefont{C.}~\bibnamefont{Weedbrook}},
  \bibinfo{author}{\bibfnamefont{T.~C.} \bibnamefont{Ralph}}, \bibnamefont{and}
  \bibinfo{author}{\bibfnamefont{P.~K.} \bibnamefont{Lam}},
  \bibinfo{journal}{Phys. Rev. Lett.} \textbf{\bibinfo{volume}{95}},
  \bibinfo{pages}{180503} (\bibinfo{year}{2005}).

\bibitem[{\citenamefont{Huang et~al.}(2015)\citenamefont{Huang, Huang, Lin,
  Wang, and Zeng}}]{huang2015high}
\bibinfo{author}{\bibfnamefont{D.}~\bibnamefont{Huang}},
  \bibinfo{author}{\bibfnamefont{P.}~\bibnamefont{Huang}},
  \bibinfo{author}{\bibfnamefont{D.}~\bibnamefont{Lin}},
  \bibinfo{author}{\bibfnamefont{C.}~\bibnamefont{Wang}}, \bibnamefont{and}
  \bibinfo{author}{\bibfnamefont{G.}~\bibnamefont{Zeng}},
  \bibinfo{journal}{Opt. Lett.} \textbf{\bibinfo{volume}{40}},
  \bibinfo{pages}{3695} (\bibinfo{year}{2015}).

\bibitem[{\citenamefont{Yin et~al.}(2019)\citenamefont{Yin, Zhu, and
  Fu}}]{yin2019phase}
\bibinfo{author}{\bibfnamefont{H.-L.} \bibnamefont{Yin}},
  \bibinfo{author}{\bibfnamefont{W.}~\bibnamefont{Zhu}}, \bibnamefont{and}
  \bibinfo{author}{\bibfnamefont{Y.}~\bibnamefont{Fu}}, \bibinfo{journal}{Sci.
  Rep.} \textbf{\bibinfo{volume}{9}}, \bibinfo{pages}{49}
  (\bibinfo{year}{2019}).

\bibitem[{\citenamefont{Fossier et~al.}(2009)\citenamefont{Fossier, Diamanti,
  Debuisschert, Villing, Tualle-Brouri, and Grangier}}]{fossier2009field}
\bibinfo{author}{\bibfnamefont{S.}~\bibnamefont{Fossier}},
  \bibinfo{author}{\bibfnamefont{E.}~\bibnamefont{Diamanti}},
  \bibinfo{author}{\bibfnamefont{T.}~\bibnamefont{Debuisschert}},
  \bibinfo{author}{\bibfnamefont{A.}~\bibnamefont{Villing}},
  \bibinfo{author}{\bibfnamefont{R.}~\bibnamefont{Tualle-Brouri}},
  \bibnamefont{and} \bibinfo{author}{\bibfnamefont{P.}~\bibnamefont{Grangier}},
  \bibinfo{journal}{New J. Phys.} \textbf{\bibinfo{volume}{11}},
  \bibinfo{pages}{045023} (\bibinfo{year}{2009}).

\bibitem[{\citenamefont{Huang et~al.}(2016)\citenamefont{Huang, Huang, Li,
  Wang, Zhou, and Zeng}}]{huang2016field}
\bibinfo{author}{\bibfnamefont{D.}~\bibnamefont{Huang}},
  \bibinfo{author}{\bibfnamefont{P.}~\bibnamefont{Huang}},
  \bibinfo{author}{\bibfnamefont{H.}~\bibnamefont{Li}},
  \bibinfo{author}{\bibfnamefont{T.}~\bibnamefont{Wang}},
  \bibinfo{author}{\bibfnamefont{Y.}~\bibnamefont{Zhou}}, \bibnamefont{and}
  \bibinfo{author}{\bibfnamefont{G.}~\bibnamefont{Zeng}},
  \bibinfo{journal}{Opt. Lett.} \textbf{\bibinfo{volume}{41}},
  \bibinfo{pages}{3511} (\bibinfo{year}{2016}).

\bibitem[{\citenamefont{Jouguet et~al.}(2012)\citenamefont{Jouguet,
  Kunz-Jacques, Debuisschert, Fossier, Diamanti, All{\'e}aume, Tualle-Brouri,
  Grangier, Leverrier, Pache et~al.}}]{jouguet2012field}
\bibinfo{author}{\bibfnamefont{P.}~\bibnamefont{Jouguet}},
  \bibinfo{author}{\bibfnamefont{S.}~\bibnamefont{Kunz-Jacques}},
  \bibinfo{author}{\bibfnamefont{T.}~\bibnamefont{Debuisschert}},
  \bibinfo{author}{\bibfnamefont{S.}~\bibnamefont{Fossier}},
  \bibinfo{author}{\bibfnamefont{E.}~\bibnamefont{Diamanti}},
  \bibinfo{author}{\bibfnamefont{R.}~\bibnamefont{All{\'e}aume}},
  \bibinfo{author}{\bibfnamefont{R.}~\bibnamefont{Tualle-Brouri}},
  \bibinfo{author}{\bibfnamefont{P.}~\bibnamefont{Grangier}},
  \bibinfo{author}{\bibfnamefont{A.}~\bibnamefont{Leverrier}},
  \bibinfo{author}{\bibfnamefont{P.}~\bibnamefont{Pache}},
  \bibnamefont{et~al.}, \bibinfo{journal}{Opt. Express}
  \textbf{\bibinfo{volume}{20}}, \bibinfo{pages}{14030} (\bibinfo{year}{2012}).

\bibitem[{\citenamefont{Pirandola et~al.}(2020)\citenamefont{Pirandola,
  Andersen, Banchi, Berta, Bunandar, Colbeck, Englund, Gehring, Lupo, Ottaviani
  et~al.}}]{pirandola2020advances}
\bibinfo{author}{\bibfnamefont{S.}~\bibnamefont{Pirandola}},
  \bibinfo{author}{\bibfnamefont{U.~L.} \bibnamefont{Andersen}},
  \bibinfo{author}{\bibfnamefont{L.}~\bibnamefont{Banchi}},
  \bibinfo{author}{\bibfnamefont{M.}~\bibnamefont{Berta}},
  \bibinfo{author}{\bibfnamefont{D.}~\bibnamefont{Bunandar}},
  \bibinfo{author}{\bibfnamefont{R.}~\bibnamefont{Colbeck}},
  \bibinfo{author}{\bibfnamefont{D.}~\bibnamefont{Englund}},
  \bibinfo{author}{\bibfnamefont{T.}~\bibnamefont{Gehring}},
  \bibinfo{author}{\bibfnamefont{C.}~\bibnamefont{Lupo}},
  \bibinfo{author}{\bibfnamefont{C.}~\bibnamefont{Ottaviani}},
  \bibnamefont{et~al.}, \bibinfo{journal}{Advances in Optics and Photonics}
  \textbf{\bibinfo{volume}{12}}, \bibinfo{pages}{1012} (\bibinfo{year}{2020}).

\bibitem[{\citenamefont{Jin et~al.}(2021)\citenamefont{Jin, Guo, Wang, Li, and
  Huang}}]{jin2021key}
\bibinfo{author}{\bibfnamefont{D.}~\bibnamefont{Jin}},
  \bibinfo{author}{\bibfnamefont{Y.}~\bibnamefont{Guo}},
  \bibinfo{author}{\bibfnamefont{Y.}~\bibnamefont{Wang}},
  \bibinfo{author}{\bibfnamefont{Y.}~\bibnamefont{Li}}, \bibnamefont{and}
  \bibinfo{author}{\bibfnamefont{D.}~\bibnamefont{Huang}},
  \bibinfo{journal}{Phys. Rev. A} \textbf{\bibinfo{volume}{104}},
  \bibinfo{pages}{012616} (\bibinfo{year}{2021}).

\bibitem[{\citenamefont{Liao et~al.}(2020)\citenamefont{Liao, Xiao, Zhong, and
  Guo}}]{liao2020multi}
\bibinfo{author}{\bibfnamefont{Q.}~\bibnamefont{Liao}},
  \bibinfo{author}{\bibfnamefont{G.}~\bibnamefont{Xiao}},
  \bibinfo{author}{\bibfnamefont{H.}~\bibnamefont{Zhong}}, \bibnamefont{and}
  \bibinfo{author}{\bibfnamefont{Y.}~\bibnamefont{Guo}}, \bibinfo{journal}{New
  J. Phys.} \textbf{\bibinfo{volume}{22}}, \bibinfo{pages}{083086}
  (\bibinfo{year}{2020}).

\bibitem[{\citenamefont{Liu et~al.}(2018)\citenamefont{Liu, Huang, Peng, Fan,
  and Zeng}}]{liu2018integrating}
\bibinfo{author}{\bibfnamefont{W.}~\bibnamefont{Liu}},
  \bibinfo{author}{\bibfnamefont{P.}~\bibnamefont{Huang}},
  \bibinfo{author}{\bibfnamefont{J.}~\bibnamefont{Peng}},
  \bibinfo{author}{\bibfnamefont{J.}~\bibnamefont{Fan}}, \bibnamefont{and}
  \bibinfo{author}{\bibfnamefont{G.}~\bibnamefont{Zeng}},
  \bibinfo{journal}{Phys. Rev. A} \textbf{\bibinfo{volume}{97}},
  \bibinfo{pages}{022316} (\bibinfo{year}{2018}).

\bibitem[{\citenamefont{Mao et~al.}(2020)\citenamefont{Mao, Huang, Zhong, Wang,
  Qin, Guo, and Huang}}]{mao2020detecting}
\bibinfo{author}{\bibfnamefont{Y.}~\bibnamefont{Mao}},
  \bibinfo{author}{\bibfnamefont{W.}~\bibnamefont{Huang}},
  \bibinfo{author}{\bibfnamefont{H.}~\bibnamefont{Zhong}},
  \bibinfo{author}{\bibfnamefont{Y.}~\bibnamefont{Wang}},
  \bibinfo{author}{\bibfnamefont{H.}~\bibnamefont{Qin}},
  \bibinfo{author}{\bibfnamefont{Y.}~\bibnamefont{Guo}}, \bibnamefont{and}
  \bibinfo{author}{\bibfnamefont{D.}~\bibnamefont{Huang}},
  \bibinfo{journal}{New J. Phys.} \textbf{\bibinfo{volume}{22}},
  \bibinfo{pages}{083073} (\bibinfo{year}{2020}).

\bibitem[{\citenamefont{Xu et~al.}(2020)\citenamefont{Xu, Ma, Zhang, Lo, and
  Pan}}]{xu2020secure}
\bibinfo{author}{\bibfnamefont{F.}~\bibnamefont{Xu}},
  \bibinfo{author}{\bibfnamefont{X.}~\bibnamefont{Ma}},
  \bibinfo{author}{\bibfnamefont{Q.}~\bibnamefont{Zhang}},
  \bibinfo{author}{\bibfnamefont{H.-K.} \bibnamefont{Lo}}, \bibnamefont{and}
  \bibinfo{author}{\bibfnamefont{J.-W.} \bibnamefont{Pan}},
  \bibinfo{journal}{Rev. Mod. Phys.} \textbf{\bibinfo{volume}{92}},
  \bibinfo{pages}{025002} (\bibinfo{year}{2020}).

\bibitem[{\citenamefont{Leverrier and
  Grangier}(2009)}]{leverrier2009unconditional}
\bibinfo{author}{\bibfnamefont{A.}~\bibnamefont{Leverrier}} \bibnamefont{and}
  \bibinfo{author}{\bibfnamefont{P.}~\bibnamefont{Grangier}},
  \bibinfo{journal}{Phys. Rev. Lett.} \textbf{\bibinfo{volume}{102}},
  \bibinfo{pages}{180504} (\bibinfo{year}{2009}).

\bibitem[{\citenamefont{Leverrier and
  Grangier}(2011)}]{leverrier2011continuous}
\bibinfo{author}{\bibfnamefont{A.}~\bibnamefont{Leverrier}} \bibnamefont{and}
  \bibinfo{author}{\bibfnamefont{P.}~\bibnamefont{Grangier}},
  \bibinfo{journal}{Phys. Rev. A} \textbf{\bibinfo{volume}{83}},
  \bibinfo{pages}{042312} (\bibinfo{year}{2011}).

\bibitem[{\citenamefont{Zhao et~al.}(2009)\citenamefont{Zhao, Heid, Rigas, and
  L{\"u}tkenhaus}}]{zhao2009asymptotic}
\bibinfo{author}{\bibfnamefont{Y.-B.} \bibnamefont{Zhao}},
  \bibinfo{author}{\bibfnamefont{M.}~\bibnamefont{Heid}},
  \bibinfo{author}{\bibfnamefont{J.}~\bibnamefont{Rigas}}, \bibnamefont{and}
  \bibinfo{author}{\bibfnamefont{N.}~\bibnamefont{L{\"u}tkenhaus}},
  \bibinfo{journal}{Phys. Rev. A} \textbf{\bibinfo{volume}{79}},
  \bibinfo{pages}{012307} (\bibinfo{year}{2009}).

\bibitem[{\citenamefont{Simon}(2017)}]{simon2017towards}
\bibinfo{author}{\bibfnamefont{C.}~\bibnamefont{Simon}}, \bibinfo{journal}{Nat.
  Photonics} \textbf{\bibinfo{volume}{11}}, \bibinfo{pages}{678}
  (\bibinfo{year}{2017}).

\bibitem[{\citenamefont{Coles et~al.}(2016)\citenamefont{Coles, Metodiev, and
  L{\"u}tkenhaus}}]{coles2016numerical}
\bibinfo{author}{\bibfnamefont{P.~J.} \bibnamefont{Coles}},
  \bibinfo{author}{\bibfnamefont{E.~M.} \bibnamefont{Metodiev}},
  \bibnamefont{and}
  \bibinfo{author}{\bibfnamefont{N.}~\bibnamefont{L{\"u}tkenhaus}},
  \bibinfo{journal}{Nat. Commun.} \textbf{\bibinfo{volume}{7}},
  \bibinfo{pages}{1} (\bibinfo{year}{2016}).

\bibitem[{\citenamefont{Lin et~al.}(2019)\citenamefont{Lin, Upadhyaya, and
  L{\"u}tkenhaus}}]{lin2019asymptotic}
\bibinfo{author}{\bibfnamefont{J.}~\bibnamefont{Lin}},
  \bibinfo{author}{\bibfnamefont{T.}~\bibnamefont{Upadhyaya}},
  \bibnamefont{and}
  \bibinfo{author}{\bibfnamefont{N.}~\bibnamefont{L{\"u}tkenhaus}},
  \bibinfo{journal}{Phys. Rev. X} \textbf{\bibinfo{volume}{9}},
  \bibinfo{pages}{041064} (\bibinfo{year}{2019}).

\bibitem[{\citenamefont{Winick et~al.}(2018)\citenamefont{Winick,
  L{\"u}tkenhaus, and Coles}}]{winick2018reliable}
\bibinfo{author}{\bibfnamefont{A.}~\bibnamefont{Winick}},
  \bibinfo{author}{\bibfnamefont{N.}~\bibnamefont{L{\"u}tkenhaus}},
  \bibnamefont{and} \bibinfo{author}{\bibfnamefont{P.~J.} \bibnamefont{Coles}},
  \bibinfo{journal}{Quantum} \textbf{\bibinfo{volume}{2}}, \bibinfo{pages}{77}
  (\bibinfo{year}{2018}).

\bibitem[{\citenamefont{Zhou et~al.}(2021)\citenamefont{Zhou, Liu, Liu, Li,
  Bai, Xue, Fu, Yin, and Chen}}]{zhou2021machine}
\bibinfo{author}{\bibfnamefont{M.-G.} \bibnamefont{Zhou}},
  \bibinfo{author}{\bibfnamefont{Z.-P.} \bibnamefont{Liu}},
  \bibinfo{author}{\bibfnamefont{W.-B.} \bibnamefont{Liu}},
  \bibinfo{author}{\bibfnamefont{C.-L.} \bibnamefont{Li}},
  \bibinfo{author}{\bibfnamefont{J.-L.} \bibnamefont{Bai}},
  \bibinfo{author}{\bibfnamefont{Y.-R.} \bibnamefont{Xue}},
  \bibinfo{author}{\bibfnamefont{Y.}~\bibnamefont{Fu}},
  \bibinfo{author}{\bibfnamefont{H.-L.} \bibnamefont{Yin}}, \bibnamefont{and}
  \bibinfo{author}{\bibfnamefont{Z.-B.} \bibnamefont{Chen}},
  \bibinfo{journal}{arXiv preprint arXiv:2108.02578}  (\bibinfo{year}{2021}).

\bibitem[{\citenamefont{Hu et~al.}(2021)\citenamefont{Hu, Im, Lin,
  L{\"u}tkenhaus, and Wolkowicz}}]{hu2021robust}
\bibinfo{author}{\bibfnamefont{H.}~\bibnamefont{Hu}},
  \bibinfo{author}{\bibfnamefont{J.}~\bibnamefont{Im}},
  \bibinfo{author}{\bibfnamefont{J.}~\bibnamefont{Lin}},
  \bibinfo{author}{\bibfnamefont{N.}~\bibnamefont{L{\"u}tkenhaus}},
  \bibnamefont{and}
  \bibinfo{author}{\bibfnamefont{H.}~\bibnamefont{Wolkowicz}},
  \bibinfo{journal}{arXiv preprint arXiv:2104.03847}  (\bibinfo{year}{2021}).

\bibitem[{\citenamefont{Yu and Zhu}(2020)}]{yu2020hyper}
\bibinfo{author}{\bibfnamefont{T.}~\bibnamefont{Yu}} \bibnamefont{and}
  \bibinfo{author}{\bibfnamefont{H.}~\bibnamefont{Zhu}},
  \bibinfo{journal}{arXiv preprint arXiv:2003.05689}  (\bibinfo{year}{2020}).

\bibitem[{\citenamefont{Shahriari et~al.}(2015)\citenamefont{Shahriari,
  Swersky, Wang, Adams, and De~Freitas}}]{shahriari2015taking}
\bibinfo{author}{\bibfnamefont{B.}~\bibnamefont{Shahriari}},
  \bibinfo{author}{\bibfnamefont{K.}~\bibnamefont{Swersky}},
  \bibinfo{author}{\bibfnamefont{Z.}~\bibnamefont{Wang}},
  \bibinfo{author}{\bibfnamefont{R.~P.} \bibnamefont{Adams}}, \bibnamefont{and}
  \bibinfo{author}{\bibfnamefont{N.}~\bibnamefont{De~Freitas}},
  \bibinfo{journal}{Proceedings of the IEEE} \textbf{\bibinfo{volume}{104}},
  \bibinfo{pages}{148} (\bibinfo{year}{2015}).

\bibitem[{\citenamefont{Liu et~al.}(2021)\citenamefont{Liu, Li, Xie, Weng, Gu,
  Cao, Lu, Li, Yin, and Chen}}]{PRXQuantum.2.040334}
\bibinfo{author}{\bibfnamefont{W.-B.} \bibnamefont{Liu}},
  \bibinfo{author}{\bibfnamefont{C.-L.} \bibnamefont{Li}},
  \bibinfo{author}{\bibfnamefont{Y.-M.} \bibnamefont{Xie}},
  \bibinfo{author}{\bibfnamefont{C.-X.} \bibnamefont{Weng}},
  \bibinfo{author}{\bibfnamefont{J.}~\bibnamefont{Gu}},
  \bibinfo{author}{\bibfnamefont{X.-Y.} \bibnamefont{Cao}},
  \bibinfo{author}{\bibfnamefont{Y.-S.} \bibnamefont{Lu}},
  \bibinfo{author}{\bibfnamefont{B.-H.} \bibnamefont{Li}},
  \bibinfo{author}{\bibfnamefont{H.-L.} \bibnamefont{Yin}}, \bibnamefont{and}
  \bibinfo{author}{\bibfnamefont{Z.-B.} \bibnamefont{Chen}},
  \bibinfo{journal}{PRX Quantum} \textbf{\bibinfo{volume}{2}},
  \bibinfo{pages}{040334} (\bibinfo{year}{2021}).

\bibitem[{\citenamefont{Devetak and Winter}(2005)}]{devetak2005distillation}
\bibinfo{author}{\bibfnamefont{I.}~\bibnamefont{Devetak}} \bibnamefont{and}
  \bibinfo{author}{\bibfnamefont{A.}~\bibnamefont{Winter}},
  \bibinfo{journal}{Proceedings of the Royal Society A: Mathematical, Physical
  and engineering sciences} \textbf{\bibinfo{volume}{461}},
  \bibinfo{pages}{207} (\bibinfo{year}{2005}).

\bibitem[{\citenamefont{Ghorai et~al.}(2019)\citenamefont{Ghorai, Grangier,
  Diamanti, and Leverrier}}]{ghorai2019asymptotic}
\bibinfo{author}{\bibfnamefont{S.}~\bibnamefont{Ghorai}},
  \bibinfo{author}{\bibfnamefont{P.}~\bibnamefont{Grangier}},
  \bibinfo{author}{\bibfnamefont{E.}~\bibnamefont{Diamanti}}, \bibnamefont{and}
  \bibinfo{author}{\bibfnamefont{A.}~\bibnamefont{Leverrier}},
  \bibinfo{journal}{Phys. Rev. X} \textbf{\bibinfo{volume}{9}},
  \bibinfo{pages}{021059} (\bibinfo{year}{2019}).

\bibitem[{\citenamefont{Bergstra et~al.}(2011)\citenamefont{Bergstra, Bardenet,
  Bengio, and K{\'e}gl}}]{bergstra2011algorithms}
\bibinfo{author}{\bibfnamefont{J.}~\bibnamefont{Bergstra}},
  \bibinfo{author}{\bibfnamefont{R.}~\bibnamefont{Bardenet}},
  \bibinfo{author}{\bibfnamefont{Y.}~\bibnamefont{Bengio}}, \bibnamefont{and}
  \bibinfo{author}{\bibfnamefont{B.}~\bibnamefont{K{\'e}gl}},
  \bibinfo{journal}{Advances in neural information processing systems}
  \textbf{\bibinfo{volume}{24}} (\bibinfo{year}{2011}).

\bibitem[{\citenamefont{Bergstra and Bengio}(2012)}]{bergstra2012random}
\bibinfo{author}{\bibfnamefont{J.}~\bibnamefont{Bergstra}} \bibnamefont{and}
  \bibinfo{author}{\bibfnamefont{Y.}~\bibnamefont{Bengio}},
  \bibinfo{journal}{Journal of machine learning research}
  \textbf{\bibinfo{volume}{13}} (\bibinfo{year}{2012}).

\bibitem[{\citenamefont{Williams and Rasmussen}(2006)}]{williams2006gaussian}
\bibinfo{author}{\bibfnamefont{C.~K.} \bibnamefont{Williams}} \bibnamefont{and}
  \bibinfo{author}{\bibfnamefont{C.~E.} \bibnamefont{Rasmussen}},
  \emph{\bibinfo{title}{Gaussian processes for machine learning}},
  vol.~\bibinfo{volume}{2} (\bibinfo{publisher}{MIT press Cambridge, MA},
  \bibinfo{year}{2006}).

\bibitem[{\citenamefont{Breiman}(2001)}]{breiman2001random}
\bibinfo{author}{\bibfnamefont{L.}~\bibnamefont{Breiman}},
  \bibinfo{journal}{Machine learning} \textbf{\bibinfo{volume}{45}},
  \bibinfo{pages}{5} (\bibinfo{year}{2001}).

\bibitem[{\citenamefont{Hutter et~al.}(2015)\citenamefont{Hutter, L{\"u}cke,
  and Schmidt-Thieme}}]{hutter2015beyond}
\bibinfo{author}{\bibfnamefont{F.}~\bibnamefont{Hutter}},
  \bibinfo{author}{\bibfnamefont{J.}~\bibnamefont{L{\"u}cke}},
  \bibnamefont{and}
  \bibinfo{author}{\bibfnamefont{L.}~\bibnamefont{Schmidt-Thieme}},
  \bibinfo{journal}{KI-K{\"u}nstliche Intelligenz}
  \textbf{\bibinfo{volume}{29}}, \bibinfo{pages}{329} (\bibinfo{year}{2015}).

\bibitem[{\citenamefont{Bergstra et~al.}(2013)\citenamefont{Bergstra, Yamins,
  and Cox}}]{bergstra2013hyperopt}
\bibinfo{author}{\bibfnamefont{J.}~\bibnamefont{Bergstra}},
  \bibinfo{author}{\bibfnamefont{D.}~\bibnamefont{Yamins}}, \bibnamefont{and}
  \bibinfo{author}{\bibfnamefont{D.~D.} \bibnamefont{Cox}}, in
  \emph{\bibinfo{booktitle}{Proceedings of the 12th Python in science
  conference}} (\bibinfo{organization}{Citeseer}, \bibinfo{year}{2013}),
  vol.~\bibinfo{volume}{13}, p.~\bibinfo{pages}{20}.

\bibitem[{\citenamefont{Srivastava et~al.}(2014)\citenamefont{Srivastava,
  Hinton, Krizhevsky, Sutskever, and Salakhutdinov}}]{srivastava2014dropout}
\bibinfo{author}{\bibfnamefont{N.}~\bibnamefont{Srivastava}},
  \bibinfo{author}{\bibfnamefont{G.}~\bibnamefont{Hinton}},
  \bibinfo{author}{\bibfnamefont{A.}~\bibnamefont{Krizhevsky}},
  \bibinfo{author}{\bibfnamefont{I.}~\bibnamefont{Sutskever}},
  \bibnamefont{and}
  \bibinfo{author}{\bibfnamefont{R.}~\bibnamefont{Salakhutdinov}},
  \bibinfo{journal}{The journal of machine learning research}
  \textbf{\bibinfo{volume}{15}}, \bibinfo{pages}{1929} (\bibinfo{year}{2014}).

\bibitem[{\citenamefont{Hornik et~al.}(1989)\citenamefont{Hornik, Stinchcombe,
  and White}}]{hornik1989multilayer}
\bibinfo{author}{\bibfnamefont{K.}~\bibnamefont{Hornik}},
  \bibinfo{author}{\bibfnamefont{M.}~\bibnamefont{Stinchcombe}},
  \bibnamefont{and} \bibinfo{author}{\bibfnamefont{H.}~\bibnamefont{White}},
  \bibinfo{journal}{Neural networks} \textbf{\bibinfo{volume}{2}},
  \bibinfo{pages}{359} (\bibinfo{year}{1989}).

\bibitem[{\citenamefont{Kingma and Ba}(2014)}]{kingma2014adam}
\bibinfo{author}{\bibfnamefont{D.~P.} \bibnamefont{Kingma}} \bibnamefont{and}
  \bibinfo{author}{\bibfnamefont{J.}~\bibnamefont{Ba}}, \bibinfo{journal}{arXiv
  preprint arXiv:1412.6980}  (\bibinfo{year}{2014}).

\bibitem[{\citenamefont{Wang and Lo}(2019)}]{wang2019machine}
\bibinfo{author}{\bibfnamefont{W.}~\bibnamefont{Wang}} \bibnamefont{and}
  \bibinfo{author}{\bibfnamefont{H.-K.} \bibnamefont{Lo}},
  \bibinfo{journal}{Phys. Rev. A} \textbf{\bibinfo{volume}{100}},
  \bibinfo{pages}{062334} (\bibinfo{year}{2019}).

\bibitem[{\citenamefont{M{\'e}len et~al.}(2017)\citenamefont{M{\'e}len,
  Freiwang, Luhn, Vogl, Rau, Sonnleitner, Rosenfeld, and
  Weinfurter}}]{melen2017handheld}
\bibinfo{author}{\bibfnamefont{G.}~\bibnamefont{M{\'e}len}},
  \bibinfo{author}{\bibfnamefont{P.}~\bibnamefont{Freiwang}},
  \bibinfo{author}{\bibfnamefont{J.}~\bibnamefont{Luhn}},
  \bibinfo{author}{\bibfnamefont{T.}~\bibnamefont{Vogl}},
  \bibinfo{author}{\bibfnamefont{M.}~\bibnamefont{Rau}},
  \bibinfo{author}{\bibfnamefont{C.}~\bibnamefont{Sonnleitner}},
  \bibinfo{author}{\bibfnamefont{W.}~\bibnamefont{Rosenfeld}},
  \bibnamefont{and}
  \bibinfo{author}{\bibfnamefont{H.}~\bibnamefont{Weinfurter}}, in
  \emph{\bibinfo{booktitle}{Quantum Information and Measurement}}
  (\bibinfo{organization}{Optical Society of America}, \bibinfo{year}{2017}),
  pp. \bibinfo{pages}{QT6A--57}.

\bibitem[{\citenamefont{Hill et~al.}(2017)\citenamefont{Hill, Chapman, Herndon,
  Chopp, Gauthier, and Kwiat}}]{hill2017drone}
\bibinfo{author}{\bibfnamefont{A.~D.} \bibnamefont{Hill}},
  \bibinfo{author}{\bibfnamefont{J.}~\bibnamefont{Chapman}},
  \bibinfo{author}{\bibfnamefont{K.}~\bibnamefont{Herndon}},
  \bibinfo{author}{\bibfnamefont{C.}~\bibnamefont{Chopp}},
  \bibinfo{author}{\bibfnamefont{D.~J.} \bibnamefont{Gauthier}},
  \bibnamefont{and} \bibinfo{author}{\bibfnamefont{P.}~\bibnamefont{Kwiat}},
  \bibinfo{journal}{Urbana} \textbf{\bibinfo{volume}{51}},
  \bibinfo{pages}{61801} (\bibinfo{year}{2017}).

\bibitem[{\citenamefont{Liao et~al.}(2017)\citenamefont{Liao, Cai, Liu, Zhang,
  Li, Ren, Yin, Shen, Cao, Li et~al.}}]{liao2017satellite}
\bibinfo{author}{\bibfnamefont{S.-K.} \bibnamefont{Liao}},
  \bibinfo{author}{\bibfnamefont{W.-Q.} \bibnamefont{Cai}},
  \bibinfo{author}{\bibfnamefont{W.-Y.} \bibnamefont{Liu}},
  \bibinfo{author}{\bibfnamefont{L.}~\bibnamefont{Zhang}},
  \bibinfo{author}{\bibfnamefont{Y.}~\bibnamefont{Li}},
  \bibinfo{author}{\bibfnamefont{J.-G.} \bibnamefont{Ren}},
  \bibinfo{author}{\bibfnamefont{J.}~\bibnamefont{Yin}},
  \bibinfo{author}{\bibfnamefont{Q.}~\bibnamefont{Shen}},
  \bibinfo{author}{\bibfnamefont{Y.}~\bibnamefont{Cao}},
  \bibinfo{author}{\bibfnamefont{Z.-P.} \bibnamefont{Li}},
  \bibnamefont{et~al.}, \bibinfo{journal}{Nature}
  \textbf{\bibinfo{volume}{549}}, \bibinfo{pages}{43} (\bibinfo{year}{2017}).

\bibitem[{\citenamefont{Lu et~al.}(2019)\citenamefont{Lu, Yin, Wang, Cui, Teng,
  Wang, Chen, Huang, Xu, Guo et~al.}}]{lu2019parameter}
\bibinfo{author}{\bibfnamefont{F.-Y.} \bibnamefont{Lu}},
  \bibinfo{author}{\bibfnamefont{Z.-Q.} \bibnamefont{Yin}},
  \bibinfo{author}{\bibfnamefont{C.}~\bibnamefont{Wang}},
  \bibinfo{author}{\bibfnamefont{C.-H.} \bibnamefont{Cui}},
  \bibinfo{author}{\bibfnamefont{J.}~\bibnamefont{Teng}},
  \bibinfo{author}{\bibfnamefont{S.}~\bibnamefont{Wang}},
  \bibinfo{author}{\bibfnamefont{W.}~\bibnamefont{Chen}},
  \bibinfo{author}{\bibfnamefont{W.}~\bibnamefont{Huang}},
  \bibinfo{author}{\bibfnamefont{B.-J.} \bibnamefont{Xu}},
  \bibinfo{author}{\bibfnamefont{G.-C.} \bibnamefont{Guo}},
  \bibnamefont{et~al.}, \bibinfo{journal}{JOSA B}
  \textbf{\bibinfo{volume}{36}}, \bibinfo{pages}{B92} (\bibinfo{year}{2019}).

\bibitem[{\citenamefont{Ding et~al.}(2020)\citenamefont{Ding, Liu, Zhang, and
  Wang}}]{ding2020predicting}
\bibinfo{author}{\bibfnamefont{H.-J.} \bibnamefont{Ding}},
  \bibinfo{author}{\bibfnamefont{J.-Y.} \bibnamefont{Liu}},
  \bibinfo{author}{\bibfnamefont{C.-M.} \bibnamefont{Zhang}}, \bibnamefont{and}
  \bibinfo{author}{\bibfnamefont{Q.}~\bibnamefont{Wang}},
  \bibinfo{journal}{Quantum Inf. Process.} \textbf{\bibinfo{volume}{19}},
  \bibinfo{pages}{1} (\bibinfo{year}{2020}).

\bibitem[{\citenamefont{Leverrier et~al.}(2010)\citenamefont{Leverrier,
  Grosshans, and Grangier}}]{leverrier2010finite}
\bibinfo{author}{\bibfnamefont{A.}~\bibnamefont{Leverrier}},
  \bibinfo{author}{\bibfnamefont{F.}~\bibnamefont{Grosshans}},
  \bibnamefont{and} \bibinfo{author}{\bibfnamefont{P.}~\bibnamefont{Grangier}},
  \bibinfo{journal}{Phys. Rev. A} \textbf{\bibinfo{volume}{81}},
  \bibinfo{pages}{062343} (\bibinfo{year}{2010}).

\bibitem[{\citenamefont{Almeida et~al.}(2021)\citenamefont{Almeida, Pereira,
  Muga, Fac{\~a}o, Pinto, and Silva}}]{almeida2021secret}
\bibinfo{author}{\bibfnamefont{M.}~\bibnamefont{Almeida}},
  \bibinfo{author}{\bibfnamefont{D.}~\bibnamefont{Pereira}},
  \bibinfo{author}{\bibfnamefont{N.~J.} \bibnamefont{Muga}},
  \bibinfo{author}{\bibfnamefont{M.}~\bibnamefont{Fac{\~a}o}},
  \bibinfo{author}{\bibfnamefont{A.~N.} \bibnamefont{Pinto}}, \bibnamefont{and}
  \bibinfo{author}{\bibfnamefont{N.~A.} \bibnamefont{Silva}},
  \bibinfo{journal}{Opt. Express} \textbf{\bibinfo{volume}{29}},
  \bibinfo{pages}{38669} (\bibinfo{year}{2021}).

\end{thebibliography}

\end{document}